\documentclass[fleqn,usenatbib]{mnras}
\usepackage{graphicx}	
\usepackage{amsmath}	
\usepackage{amssymb}
\usepackage{subcaption}
\usepackage{multirow}
\usepackage{amsmath}
\usepackage{url}
\usepackage{setspace}
\usepackage{longtable}
\usepackage{pdflscape}
\usepackage{float}
\usepackage{booktabs}
\usepackage{tabularx}
\usepackage{amssymb}
\usepackage{wasysym}
\usepackage{rotating}
\usepackage{adjustbox}
\usepackage{xspace}
\usepackage{xcolor}
\usepackage[T1]{fontenc}
\usepackage{adjustbox}
\usepackage{enumitem,calc}
\usepackage{hyperref}
\usepackage{lscape} 

\newcommand{\hi}{{H\,{\small I}}\xspace }

\newcommand{\simba}{{\sc Simba}}

\newcommand\Tstrut{\rule{0pt}{3ex}}         

\title[MIGHTEE-\hi: Mass Models and Dark Matter properties]{MIGHTEE-\hi: Mass Models and Dark Matter properties}

\author[A.~A.~Ponomareva et al.]{Anastasia A.~Ponomareva$^{1,2}$\thanks{Email: a.ponomareva2@herts.ac.uk}, P. E. Mancera Pi$\mathrm{\tilde{n}}$a$^3$, A. A.  Vărăşteanu$^{2}$, M. Glowacki$^{4,6}$, H. Desmond$^{7}$, 
\and M. J. Jarvis$^{2,8}$, T. Yasin$^{2}$, I. Heywood$^{2,9,10}$, N. Maddox$^{11}$,
E. A. K. Adams$^{12,13}$, M. Baes$^{14}$, A. Gebek$^{14}$,
\and S. Kurapati$^{12}$, M. Maksymowicz-Maciata$^{11}$, K. A. Oman$^{15,16}$, H. Pan$^{17,2}$, I. Prandoni$^{18}$, 
\and S. H. A. Rajohnson$^{19}$, I. Ruffa$^{20}$ and K. Spekkens$^{21}$
\\
$^1$Centre for Astrophysics Research, School of Physics, Astronomy and Mathematics, University of Hertfordshire, College Lane,\\
Hatfield, AL10 9AB, UK\\
$^2$Oxford Astrophysics, Denys Wilkinson Building, University of Oxford, Keble Rd, Oxford, OX1 3RH, UK\\
$^3$Leiden Observatory, Leiden University, P.O. Box 9513, 2300 RA, Leiden, The Netherlands\\
$^4$Institute for Astronomy, University of Edinburgh, Royal Observatory, Edinburgh, EH9 3HJ, United Kingdom\\
$^6$Inter-University Institute for Data Intensive Astronomy, Department of Astronomy, University of Cape Town, Cape Town, South Africa\\
$^7$Institute of Cosmology \& Gravitation, University of Portsmouth, Dennis Sciama Building, Portsmouth, PO1 3FX, UK\\
$^8$Department of Physics and Astronomy, University of the Western Cape, Robert Sobukwe Road, Bellville 7535, South Africa\\
$^9$Centre for Radio Astronomy Techniques and Technologies, Department of Physics and Electronics, Rhodes University, PO Box 94, \\Makhanda, 6140, South Africa. \\
$^{10}$South African Radio Astronomy Observatory, 2 Fir Street, Black River Park, Observatory, Cape Town, 7925, South Africa.\\
$^{11}$School of Physics, H.H. Wills Physics Laboratory, Tyndall Avenue, University of Bristol, Bristol, BS8 1TL, UK\\
$^{12}$ASTRON, the Netherlands Institute for Radio Astronomy, Oude Hoogeveesedijk 4,7991 PD Dwingeloo, The Netherlands\\
$^{13}$Kapteyn Astronomical Institute, PO Box 800, 9700 AV Groningen, The Netherlands\\
$^{14}$ Sterrenkundig Observatorium, Universiteit Gent, Krijgslaan 281 S9, B-9000 Gent, Belgium\\
$^{15}$Institute for Computational Cosmology, Physics Department, Durham University, South Road, Durham DH1 3LE, United Kingdom\\
$^{16}$Centre for Extragalactic Astronomy, Physics Department, Durham University, South Road, Durham DH1 3LE, United Kingdom\\
$^{17}$National Astronomical Observatories, Chinese Academy of Sciences, Beijing 100101, People’s Republic of China\\
$^{18}$INAF-IRA, Via P. Gobetti 101, 40129, Italy\\
$^{19}$INAF - Osservatorio Astronomico di Cagliari, Via della Scienza 5, I-09047 Selargius (CA), Italy\\
$^{20}$INAF, Arcetri Astrophysical Observatory, Largo Enrico Fermi 5, I-50125 Florence, Italy\\
$^{21}$Department of Physics, Engineering Physics and Astronomy, Queen's University, Kingston, Ontario, K7L 3N6, Canada\\
}

\date{Accepted 2025  XX. Received 2025  XX; in original form 2025  XX}

\pubyear{\the\year{}}

\begin{document}
\label{firstpage}
\pagerange{\pageref{firstpage}--\pageref{lastpage}}
\maketitle

\begin{abstract}
Measuring galaxy rotation curves is critical for inferring the properties of dark-matter haloes in the Lambda Cold Dark Matter ($\Lambda$CDM) paradigm.
We present H\,\textsc{i} rotation curves and mass models for 20 galaxies from the MIGHTEE survey. Using extended H\,\textsc{i} kinematics, we construct resolved mass models that include stellar, gaseous, and dark-matter components. Stellar masses are derived using 3.6\,$\mu$m imaging under fixed mass-to-light ratio ($\Upsilon_{*} = M/L$) assumptions and are complemented, for the first time for a \hi-selected sample, by spatially resolved $M/L$, obtained from multi-wavelength SED fitting.
We examine the ratio of baryonic to observed rotation velocity ($V_{\rm bar}/V_{\rm obs}$) at the characteristic radius $R_{2.2}$. Adopting a fixed $\Upsilon_\star = 0.5\,M_\odot/L_\odot$ yields a clear dependence of $V_{2.2}/V_{\rm obs}$ on galaxy luminosity, while adopting $\Upsilon_\star = 0.2\,M_\odot/L_\odot$ substantially weakens this trend. In contrast, the resolved $M/L$ analysis preserves the luminosity dependence while modifying the stellar contribution on a galaxy-by-galaxy basis, providing a more accurate representation of the underlying relation.
We model the dark-matter haloes using Navarro--Frenk--White profiles and find that the different assumptions for a fixed a $M/L$ systematically shift galaxies relative to the theoretical stellar-to-halo mass and baryonic-to-halo mass relations, while the spatially varying $M/L$ yields the closest agreement with theoretical benchmarks within $\Lambda$CDM. We therefore demonstrate that future investigations of the dark matter properties of galaxies using rotation curves need to account for varying $M/L$ across individual galaxy profiles and between galaxies in order to obtain accurate measurements of the dark matter, and therefore test $\Lambda$CDM.

\end{abstract}


\section{Introduction}
\label{sec:intro}
Since the seminal discoveries in the late 20th century, which revealed that galaxy rotation curves remain flat at large radii rather than declining as predicted by Keplerian dynamics, it has been strongly suggested that most of the mass in galaxies is not visible \citep{bosma1978, rubin1978, bosma1981}. This phenomenon provided compelling evidence for the existence of dark matter \citep{vanalbada1985, begeman1991}. 
The study of galaxy rotation curves has since been a cornerstone in understanding the distribution of matter in disc galaxies, as well as the connection between baryons and the dark matter (DM) halo. Rotation curves offer direct insight into the gravitational potential of galaxies, since the observed rotation curve consists of contributions from all dynamical components: stars, gas, and DM. Therefore, studying rotation curves is critical for inferring the presence and properties of DM haloes in the Lambda Cold Dark Matter ($\Lambda$CDM) paradigm.

Mass modelling of galaxies, which decomposes the total gravitational potential into contributions from baryonic matter (stars and gas) and DM, is a crucial tool for understanding the structure and dynamics of galaxies \citep{deblok2008, martinsson13, aniyan2018, aniyan2021}. By deconstructing the observed rotation curves into the individual contributions from all dynamic components, we can explore the relative contributions of different galactic components to the overall rotation curve and hence determine the properties of the DM halo \citep{deblok2010, oh2015,readiorio2016, mancerapina2022b}. This type of mass modelling provides insights not only into the DM itself but also the processes that shape galaxy formation and evolution, such as feedback from star formation and active galactic nuclei, which can affect the distribution of both baryons and dark matter within the galaxy \citep{dicintio2016, katz2017, marasco2020,mancerapina2022b}. 

Dark matter halo models, such as the Navarro-Frenk-White (NFW) profile \citep{nfw1996} or feedback-modified profiles (e.g. DC14; \citealt{DC14} or {\sc coreNFW}; \citealt{read2016, allaert2017}), are essential for simulations and theoretical models of galaxy formation and evolution. These models provide parameterised descriptions of the DM halo structure, which can then be tested against the observed rotation curves. The NFW profile, for example, predicts a cuspy halo with a steep density gradient near the centre, as derived from dark matter-only simulations \citep{springel2005,klypin2011}. However, observed rotation curves, especially in low-mass and low-surface-brightness galaxies, often favour dark matter profiles with shallower inner slopes, implying the presence of DM cores rather than cusps \citep{deblok2008, oh2015,mancerapina2025}. The DC14 or {\sc coreNFW} model (e.g. \citealt{read2016, readiorio2016, allaert2017}), which incorporates the effects of baryonic feedback processes (such as supernova-driven outflows) on dark matter halos, provides a more accurate fit to the observed data in these cases \citep{dicintio2016, katz2017}. These empirical models are crucial for both interpreting observations and refining the theoretical frameworks of galaxy formation. Testing these models against a broad range of galaxy types and properties allows us to assess how well they capture the complexities of galaxy evolution, and whether they are consistent with $\Lambda$CDM  predictions. 

Understanding the mass distribution of galaxies within their DM halos at $z \sim 0$ is important for tracing the Universe’s assembly history. Theoretical simulations that include baryonic processes, such as gas outflows and star formation, predict that DM halo structures evolve over time \citep{DC14,tollet2016}. By comparing mass models from observed rotation curves and established DM scaling relations, e.g. the stellar‐to‐halo mass relation (SHMR) with these simulations, we can evaluate their accuracy and refine our understanding of cosmic growth \citep{katz2017,posti2019,marasco2020,diteodoro2023}.

To robustly measure these halo profiles, high‐quality rotation curves extending into the outer regions are essential. For galactic mass distribution studies, rotation curves derived from neutral atomic hydrogen (\hi) have advantages over optical or other cold gas tracers \citep{deblok2008, lelli2016, martinsson13}, because \hi discs tend to extend farther into a galaxy’s outskirts than stellar discs \citep{deblok2008, halogas}. This extensive spatial coverage allows for the mapping of rotational velocities to a larger radii, providing critical insights into the mass distribution where DM dominates the gravitational potential over baryons \citep{bosma1978}. 
Furthermore, \hi is dynamically cold (with a velocity dispersion of 
$\sigma_{v} \sim 10 \, \rm km s^{-1}$ \citep{Ianjamasimanana2015,mogotsi2016,mancerapina2025} and follows nearly circular orbits, making it ideal for reliably tracing the total gravitational potential \citep{iorio2017, pavel2021}. In contrast, optical and CO tracers are confined to the inner regions due to the much more compact distribution of stars and molecular gas, resulting in a less comprehensive view of a galaxy's mass profile, and tend to be more susceptible to the mass modelling degeneracy, such as e.g. disc-halo degeneracy, \citep{ frank2016, deblok2016, combes2019A&A...623A..79C, balmaverde2021, venturi2021}. These advantages make \hi rotation curves an indispensable tool for accurate mass modelling in galactic studies.

A significant challenge in mass modelling of observed rotation curves arises from the disc-halo degeneracy, where the contributions of the stellar disc and dark matter halo to the overall gravitational potential can be difficult to disentangle. This degeneracy arises from the difficulty to accurately determine the mass of the stellar disc, particularly since both the disc and halo influence the shape of the rotation curve \citep{bershady2010A, aniyan2018, aniyan2021}. Methods to address this issue include using stellar population synthesis models to estimate the mass-to-light ratio ($\Upsilon_{*}$) of the stellar disc \citep{schombert2019, schombert2022}, or applying the maximum or minimum disc hypotheses, which assume that the disc contributes the maximum or minimum amount of mass allowable by the observed rotation curve. Additionally, other techniques, such as measuring the vertical velocity dispersion of stars in the disc, can provide further constraints to break the degeneracy \citep{bershady2010A, martinsson13,aniyan2018,aniyan2021}. 

\begin{figure*} 
\centering
\includegraphics[scale=0.7]{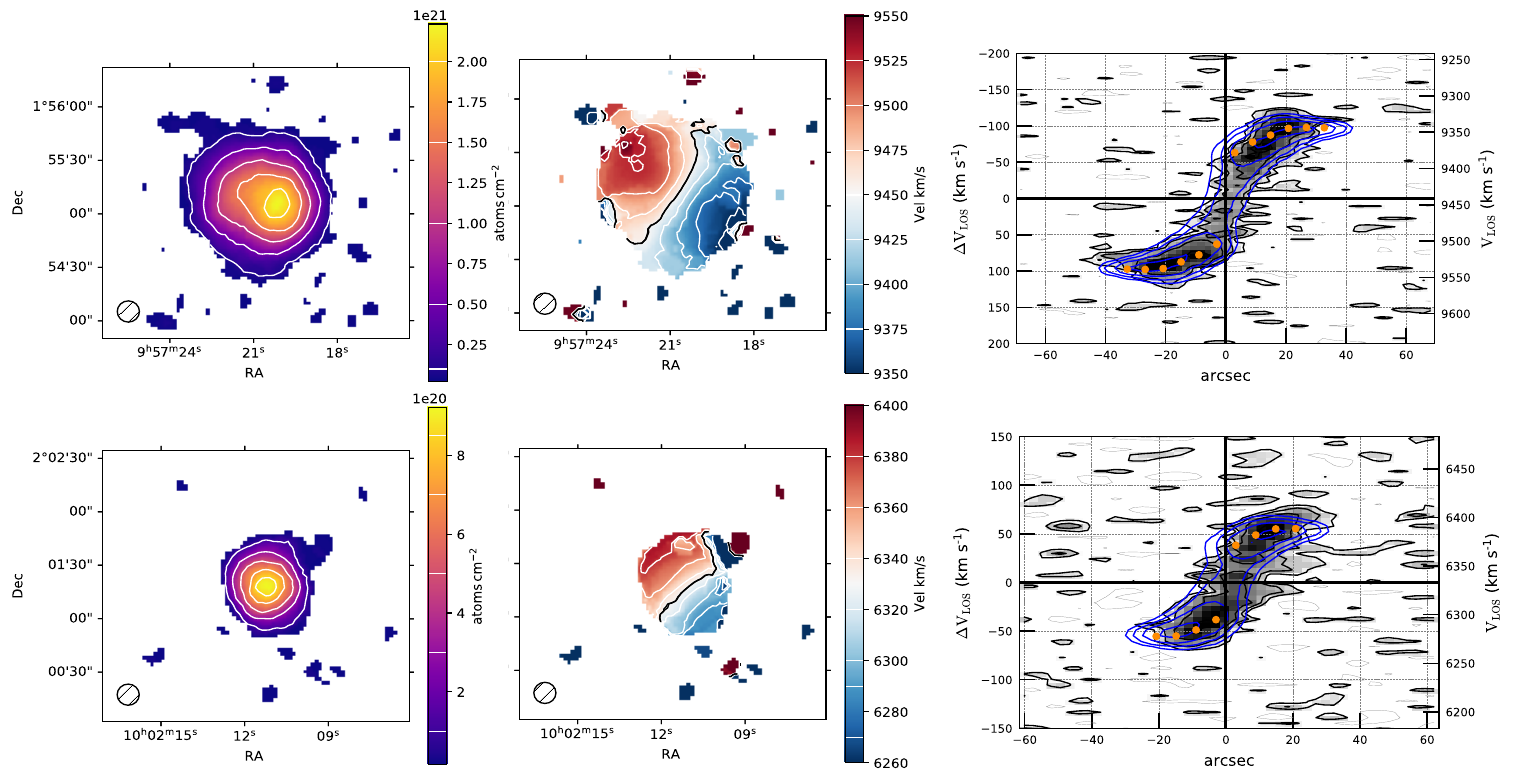}
\caption{\hi moment maps and position–velocity (PV) diagrams for two representative galaxies, J095720.6+015507 (top) and J100211.2+020118 (bottom). The left panel shows the H I moment-0 map, the central panel the moment-1 map, and the right panel the position–velocity (PV) diagram along the major axis, with the derived line-of-sight rotation curve (orange) and model (blue) overplotted. The synthesised beam is shown as an ellipse in the lower left corner.}
\label{fig_moments}
\end{figure*}

In this paper, we present resolved \hi\ rotation curves for a sample of 20 galaxies selected from the MeerKAT International GigaHertz Tiered Extragalactic Exploration (MIGHTEE) \hi Data Release (DR) 1 \citep{heywood2024} and utilising L2 band spanning a period of approximately one billion years in lookback time ($0 < z < 0.08$). By combining \hi\ observations with ancillary Spitzer photometry and resolved stellar mass surface densities from \citet{Varasteanu2025}, we construct detailed mass models to disentangle the contributions of stars, gas, and DM in each system. We investigate how different assumptions about stellar mass measurement affect the inferred baryon-to-total dynamic mass relation. We further compare the derived DM halo properties and scaling relations, such as the stellar-to-halo and baryonic-to-halo mass relations, against theoretical predictions from abundance matching and hydrodynamic cosmological simulations. 

This paper is organised as follows. Section \ref{sec:survey} describes the latest MIGHTEE-\hi\ data release and the \hi\ data utilised in this study. Section \ref{sec:HI} explains the derivation of \hi\ rotation curves and surface mass density profiles. Section \ref{sec:stellarmass} outlines the method used to obtain stellar brightness profiles. Section \ref{sec:mass_models} presents the constructed mass models and examines the impact of different stellar mass measurements. Section \ref{sec:dark_matter} discusses the properties of dark matter and associated scaling relations. Finally, the summary and conclusions, as well as avenues for future work are presented in Section \ref{sec:summary}.

Throughout the paper we assume a $\Lambda$CDM cosmology parameters of $H_{0}=67.4$\,km\,s$^{-1}{\rm Mpc}^{-1}$, $\Omega_{\rm m}=0.315$ and $\Omega_{\Lambda}=0.685$ \citep{planck2020}, and adopt standard values for the critical density ($\rho_{\rm crit}= 1.36 \times 10^{11} \, M_\odot \, \text{kpc}^{-3}$) and gravitational constant ($G=4.3009 \times 10^{-6} \, \text{kpc} \, M_\odot^{-1} \, (\text{km/s})^2$).


\section{HI Data}
\label{sec:survey}
MIGHTEE is a survey of four well-known deep extragalactic fields observed by MeerKAT, the SKA precursor radio interferometer located in South Africa \citep{Jonas_2009}. MeerKAT consists of 64 offset Gregorian dishes (13.5\,m diameter main reflector and 3.8\,m sub-reflector) and is equipped with three receivers: UHF-band ($580 < \nu < 1015$\,MHz), L-band ($900 < \nu < 1670$\,MHz), and S-band ($1750 < \nu < 3500$\,MHz). The MeerKAT data are collected in spectral mode, making MIGHTEE a spectral line, continuum, and polarisation survey \citep{jarvis2016}. The \hi emission project within the MIGHTEE survey (MIGHTEE--\hi) is described in detail in \citet{maddox2021}.

The Early Science (ES) MIGHTEE--\hi data have resulted in various publications and discoveries (e.g. \citealt{ranchod2021, tudorache2022}), with the source catalogue released in \citet{ponomareva2023}. For this study, we utilise the DR1 \hi data from the deep MIGHTEE spectral line observations of the COSMOS field \citep{heywood2024}. These data are deeper and cover a larger area of the COSMOS field than the ES data, with a total of 94.2\,h on-target and a close-packed mosaic of 15 individual pointings (covering $>4\,\mathrm{deg}^2$ on the sky). The spectral imaging covers two broad regions (960–1150\,MHz and 1290–1520\,MHz) within MeerKAT's L-band, with up to 26\,kHz (5,5 km\,s$^{-1}$) spectral resolution, in contrast to the ES data with coarser velocity resolution of 208 kHz (44 km\,s$^{-1}$) at $z=0$ \citep{ponomareva2021}. 

The data were processed to produce constituent images for mosaics constructed from all pointings at three different resolutions. For this study, we utilise the highest spatial resolution to ensure our galaxies are sufficiently resolved for the 3D kinematic modelling. The typical angular resolution for our data is 12 arcsec with a circular synthesised beam and a velocity resolution of 5.5 km\,s$^{-1}$ at $z=0$. The median noise in the highest spectral resolution data (5.5 km\,s$^{-1}$ at $z=0$) is 74\,$\mu$Jy\,beam$^{-1}$\,channel$^{-1}$, corresponding to a $3\sigma$ \hi column density sensitivity (N$_{\hi}$) of $9 \times 10^{18}$ cm$^{-2}$.
For a full description of the data processing, analysis, and release, see \citet{heywood2024}.

\subsection{The Sample}
For this study, we rely on the \hi source catalogue from the MIGHTEE early science (ES) data covering the COSMOS field \citep{ponomareva2023}. All 75 galaxies identified in the ES data have been recovered in the DR1 \citep{heywood2024}. At this stage we do not perform any additional source finding. 

Out of the 75 sources, we select galaxies that are resolved with at least three resolution elements across their major axes, as estimated using the $M_{\rm HI}$--$D_{\rm HI}$ relation \citep{rajohson2022}, have inclinations greater than $20^\circ$, and are detected in \hi\ with a signal-to-noise ratio exceeding 3 per channel map, ensuring reliable kinematic modelling. For a detailed description of the selection procedure for galaxies suitable for kinematic analysis, see \citealt{ponomareva2021} (Section~4.3.1).

Our final sample comprises 20 galaxies spanning a wide range of circular velocities, from $40\,\mathrm{km\,s^{-1}}$ to $200\,\mathrm{km\,s^{-1}}$, \hi masses of $10^{7} \leq M_{\mathrm{HI}}[M_{\odot}] < 10^{10}$ \citep{heywood2024}, and stellar masses of $10^{7} \leq M_{*}[M_{\odot}] < 10^{11}$ \citep{maddox2021}. The sample spans the entire redshift range of the MIGHTEE L2 band \citep{heywood2024}, covering approximately 1\,Gyr in lookback time ($0 \leq z \leq 0.08$). The \hi data (moment~0 and moment~1 maps, as well as position–velocity (PV) diagrams extracted along the major axis) for two representative galaxies from our sample are shown in Figure~\ref{fig_moments}. The main parameters of the sample used in this work can be found in Table \ref{tbl_sample}.

This sample constitutes a pilot study using the MIGHTEE–\hi DR1 in the COSMOS field, showcasing the power of the MIGHTEE survey to deliver resolved \hi kinematics in deep extragalactic fields up to $z\sim 0.1$. Future data releases covering the XMM–Large Scale Structure (XMMLSS), Extended Chandra Deep Field South (ECDFS), ELAIS-S1 and the MIGHTEE Fornax Survey (MFS) fields will extend this work to hundreds of resolved galaxies.

\section{HI rotation curves and Surface mass density profiles}
\label{sec:HI}

\begin{figure} 
\centering
\includegraphics[scale=0.65]{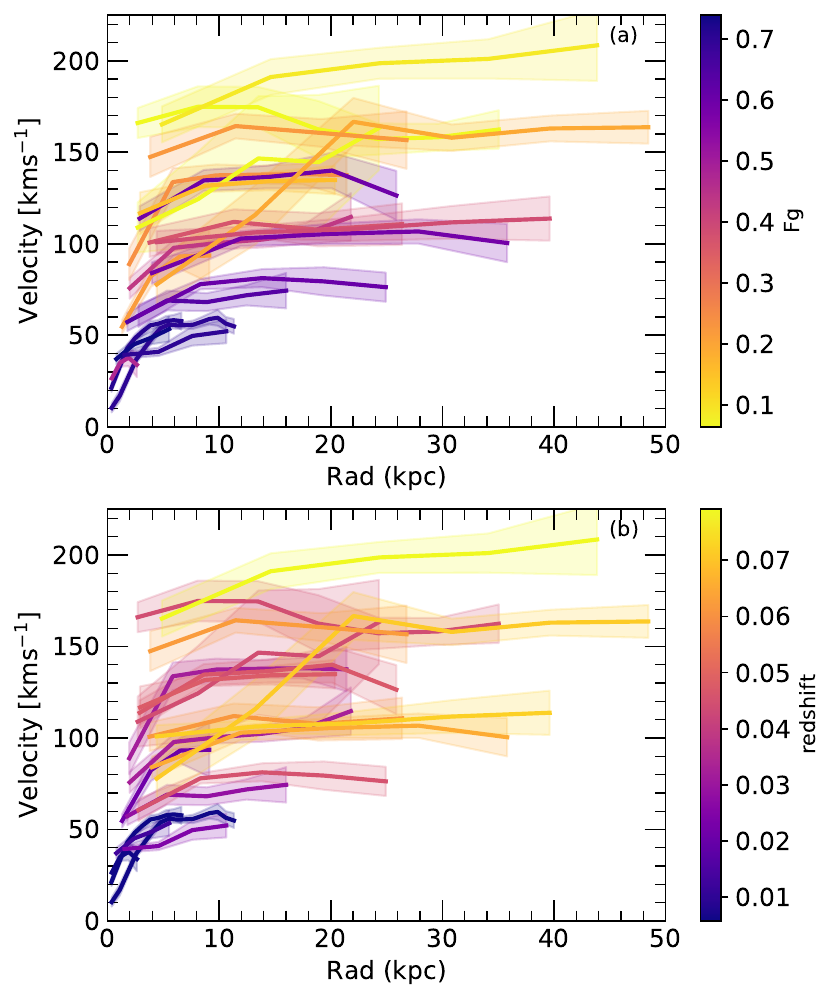}
\caption{Rotation curves of our final sample colour-coded by the galaxies gas fraction F$_{g} =M_{\mathrm{HI}}/M_{\mathrm{bar}}$ (a) and by redshift (b). The measurement uncertainties are indicated by the shaded regions}
\label{fig:rcs}
\end{figure}

\subsection{Rotation curves}
\label{sec:RCs}
Modern 3D kinematic modelling software is capable of constraining gas dynamics in galaxies resolved with as few as three resolution elements across their major axis and with SNR $\gtrsim 3$ \citep{3DBarolo, Kamphuis2015}. Moreover, unlike earlier approaches that required initial estimates of key galactic parameters, such as systemic velocity, central position, and position and inclination angles to derive a rotation curve \citep{begeman1989,mverheijen2001,ponomareva2016}, modern kinematic modelling software is able perform fitting procedures with minimal prior information.


In \citet{ponomareva2021}, we explored the capabilities of 3DBarolo \citep{3DBarolo} by fitting a sample of sources from the MIGHTEE ES data release in an automated way. Out of the original sample of 270 detections, only 68 presented reliable kinematic measurements suitable for resolved studies, leading to a study of the baryonic Tully-Fisher relation based on resolved kinematics. We have learned that while 3DBarolo is capable of fitting large samples of galaxies in an automated way without prior knowledge of any galaxy parameters, the inclination of galaxies remains a significant issue \citep{pavel2020}. The software performs much more reliably when an initial estimation of inclination angle is provided (see Figure 2 in \citealt{ponomareva2021}).

To derive resolved rotation curves for our sample of galaxies, we use inclinations measured from the Spitzer IRAC 1 band at 3.6~$\mu$m, which is widely adopted to trace the bulk of the stellar mass distribution in galaxies \citep{ponomareva2017, miguel14, meidt12}. This near-infrared band is minimally affected by dust extinction and is largely insensitive to recent star formation, providing a robust representation of the disc geometry.

The inclinations ($i$) are estimated from the axis ratios using the standard relation:
\begin{equation}
\cos^{2}(i) = \frac{(b/a)^{2} - q_{0}^{2}}{1 - q_{0}^{2}},
\label{inclination}
\end{equation}
where $b$ and $a$ are the semi-minor and semi-major axes, respectively, and $q_{0}$ represents the intrinsic axis ratio of the disc, typically ranging between 0 (infinitely thin disc) and 0.4 \citep{foque1990}. Throughout this work, we adopt a standard value of $q_{0} = 0.2$, following \citet{tully1985}.

We then use these inclinations to run 3DBarolo fully automatically, using only the data cube and the inclination as input values, keeping $\rm V_{rot} $ and $\rm V_{disp} $ as free parameters. The fully automated run adopts a ring separation equal to the beam major axis, ensuring that the kinematic measurements are largely independent and that pixel-to-pixel correlations remain negligible.\footnote{For consistency and to prepare for the large data load from the ongoing and upcoming \hi\ surveys, we run the fully automated version of 3DBarolo, following the methodology and motivation detailed in \citet{ponomareva2021}.} We set the boundary constraint for inclination fitting to $\Delta i = \pm5^\circ$ (see \citealt{ponomareva2021} for details).
 To account for the uncertainties on the derived rotation curves and \hi surface mass densities we utilise the built-in \textsc{FLAGERRORS} function in 3DBarolo, which estimates the uncertainties on the fitted parameters and the resulting rotational velocities by exploring the full parameter space, ensuring the fit converges at the minima \citep{3DBarolo}.

The resulting rotation curves for two representative galaxies from our sample are shown in Figure \ref{fig_moments} overplotted on top of the PV diagrams along the major axis. The compilation of all resulting rotation curves is shown in Figure~\ref{fig:rcs}. Given the nature of our volume‐limited, untargeted survey, it is unsurprising that we predominantly detect gas‐rich, low‐velocity galaxies at lower redshifts, whereas galaxies with higher stellar masses appear at higher redshifts (due to larger volume probed) and exhibit lower gas fractions: $\mathrm{F}{\mathrm{g}} = 1.4 M_{\mathrm{HI}} / M_{\mathrm{bar}}$, where $M_{\mathrm{bar}} = 1.4 M_{\mathrm{HI}} + M_{*}$, with the factor 1.4 accounting for the primordial abundance of helium and metals \citep{arnett1999}. 
We note that our rotation curves are not corrected for pressure support (i.e. we do not apply the asymmetric drift correction, e.g. \citealt{binney2008,readiorio2016,pavel2026}). This correction is negligible for \hi at our mass and rotational speed ranges, and becomes important only for galaxies $V_{\rm rot}/\sigma_v \lesssim 3$, which are typically dwarfs with $V_{\rm rot} \lesssim 20\,\mathrm{km\,s^{-1}}$  (e.g. \citealt{iorio2017,pavel2021}). 


 Figure~\ref{fig:rcs} (b) shows the derived rotation curves as a function of redshift. Prior to our study, resolved \hi\ rotation curve samples only covered galaxies up to redshift $z = 0.02$ \citep{deg2022,deg2024}. Although this redshift range is not expected to encompass any cosmological \hi\ evolution \citep{ponomareva2021}, it mitigates distance uncertainties arising from the local universe's cosmic flow and allows for cosmological luminosity distance measurements. The systemic velocities of our sample range from $V_{\mathrm{sys}} \approx 1700\,\mathrm{km\,s}^{-1}$ to $V_{\mathrm{sys}} \approx 22770\,\mathrm{km\,s}^{-1}$, with a mean value of $\bar{V}_{\mathrm{sys}} \approx 11800\,\mathrm{km\,s}^{-1}$, making the effect of peculiar velocities negligible for the majority of the sample \citep{tonry2000, tully2016, said2020}. 

\subsection{Gas Surface Mass Density}
\label{sec:gas_sbp}
\begin{figure} 
\centering
\includegraphics[scale=0.65]{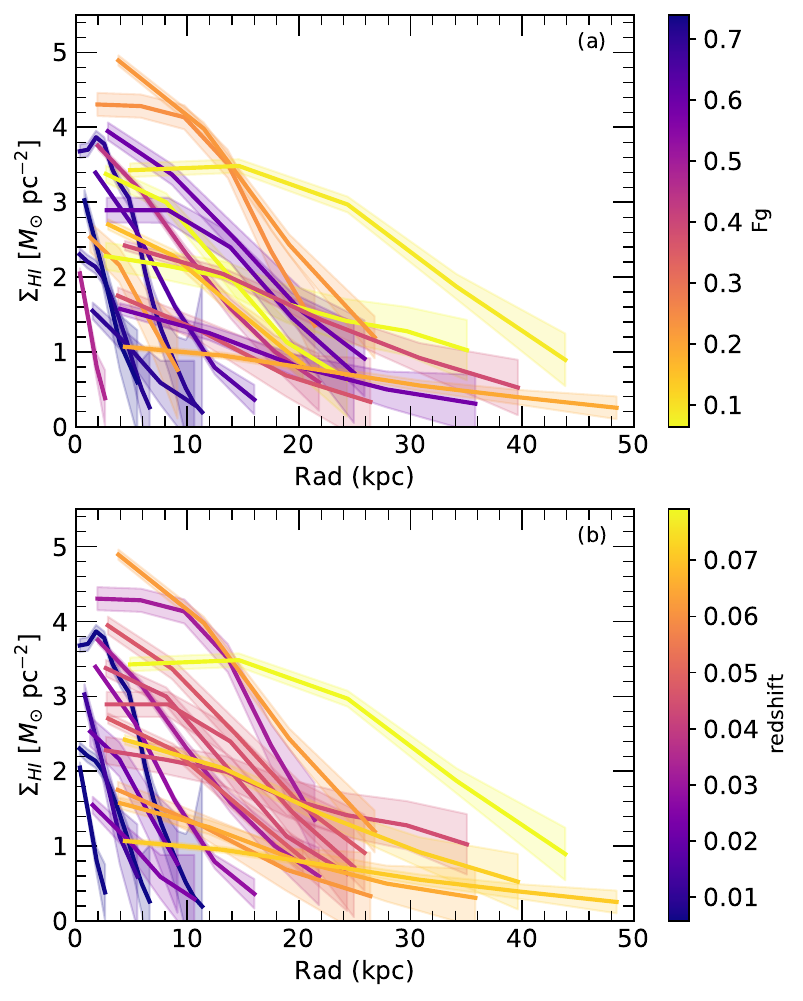}
\caption{Inclination corrected \hi surface mass density 1D profiles of our final sample, colour-coded by the galaxies' gas fraction F$_{g} =M_{\mathrm{HI}}/M_{\mathrm{bar}}$ (a) and by redshift (b). The measurement uncertainties are indicated by the shaded regions.}
\label{fig_SDPS}
\end{figure}
Following the extraction of rotation curves, we also derive resolved radial \hi\ surface mass density profiles taking advantage of 3DBarolo. At each radius, the mean \hi\ flux density is estimated from the input data cubes. We then convert these values from units of flux density to physical units of \hi\ column density, $N_{\mathrm{HI}}\, [\mathrm{atoms\, cm^{-2}}]$, following the standard procedure (see Eq.~78 in \citealt{meyer2017}). Next, we correct the column density profiles for inclination to obtain face-on values, thereby accounting for projection effects due to the galaxies' orientation. Finally, we apply the standard conversion to obtain \hi\ surface mass density, where $1\,\mathrm{M}_{\odot}\,\mathrm{pc}^{-2} = 1.249 \times 10^{20}\,\mathrm{atoms\,cm^{-2}}$. 

The resulting \hi\ surface mass density profiles are shown in Figure~\ref{fig_SDPS}. As expected, they exhibit a variety of shapes across our sample of galaxies. Higher mass galaxies tend to illustrate a central depletion of \hi\ surface density, with the neutral hydrogen gas extending out to large radii. This indicates that in these galaxies, the \hi\ gas is more distributed in the outer regions of the galactic disc. Conversely, smaller galaxies display steeper \hi\ profiles.


\section{Stellar surface brightness and mass densities}
\label{sec:stellarmass}
\subsection{Spitzer photometry}
The Spitzer IRAC 1 band at 3.6~$\mu$m is widely used to derive stellar surface brightness profiles, which are essential for subsequent mass modelling and rotation curve decomposition. One of the key advantages of the 3.6~$\mu$m wavelength is its sensitivity to the older, low-mass stars that dominate the bulk of the stellar mass in galaxies. At this wavelength, the emission is not affected by dust extinction and is predominantly due to evolved stellar populations such as red giants and main-sequence stars. This results in a more accurate tracing of the underlying stellar mass distribution compared to optical wavelengths, which can be significantly influenced by recent star formation and dust obscuration \citep{sheth2010, meidt2012}.

Furthermore, the stellar mass-to-light ratio ($\Upsilon_{*}$) in the 3.6~$\mu$m band is considered to be well understood and stable across different galaxy types and stellar populations. Studies have shown that the $\Upsilon_{*}$ at 3.6~$\mu$m exhibits minimal variation with stellar population age or metallicity \citep{rock2015}, allowing for a more direct and reliable conversion from observed luminosity to stellar mass \citep{McGaugh2014}. \citet{Meidt2014}, for example, showed that the intrinsic scatter in the $\Upsilon_{*}$ ratio at 3.6~$\mu$m is less than 0.1~dex, which is significantly lower than in optical bands. \citet{maarsco2025} performed a comparison between stellar masses inferred from 3.6~$\mu$m data and those derived through multi-wavelength SED modelling, demonstrating that SED fitting provides consistent stellar mass estimates in good agreement with dynamical estimates.

Additionally, the minimal impact of dust extinction at 3.6~$\mu$m reduces the need for uncertain extinction corrections, further improving the accuracy of the stellar mass profiles \citep{Querejeta2015}. This is particularly beneficial for edge-on galaxies or regions with high dust content, where optical observations can be severely affected. The use of 3.6~$\mu$m data thus allows for a clearer view of the stellar mass distribution, which is crucial for understanding the interplay between baryonic and dark matter in galaxies \citep{ponomareva2017,ponomareva2018}. Numerous studies have successfully utilised Spitzer IRAC 1 band data for mass modelling and rotation curve decomposition (e.g. \citealt{mcgaugh16, lelli2016, posti2019, diteodoro2023}). 

\begin{figure} 
\centering
\includegraphics[scale=0.65]{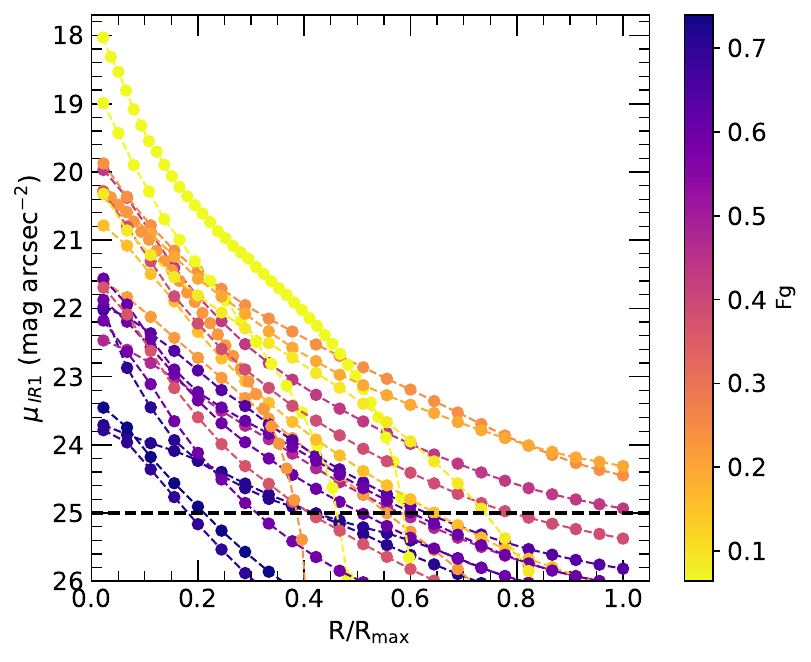}
\caption{IRAC 1 surface brightness profiles of our final sample colour-coded by the galaxies' gas fraction, $\mathrm{F_g} = M_{\mathrm{HI}} / M_{\mathrm{bar}}$ and normalized by the radius of the last measured ellipse (R$_{\rm max}$). The horizontal dashed line represents the Spitzer IRAC 1 $3\sigma$ magnitude limit}
\label{fig:sbp}
\end{figure}
Taking all of this into account, for our study we also focus on the Spitzer 3.6~$\mu$m band to derive the stellar mass surface density for the sample galaxies. To obtain the stellar surface brightness profiles necessary for mass modelling and rotation curve decomposition, we processed the galaxy images using the standard approach \citep{Varasteanu2025} and extracted the stellar surface brightness profile of each galaxy.  

\begin{figure*} 
\centering
\includegraphics[scale=0.6]{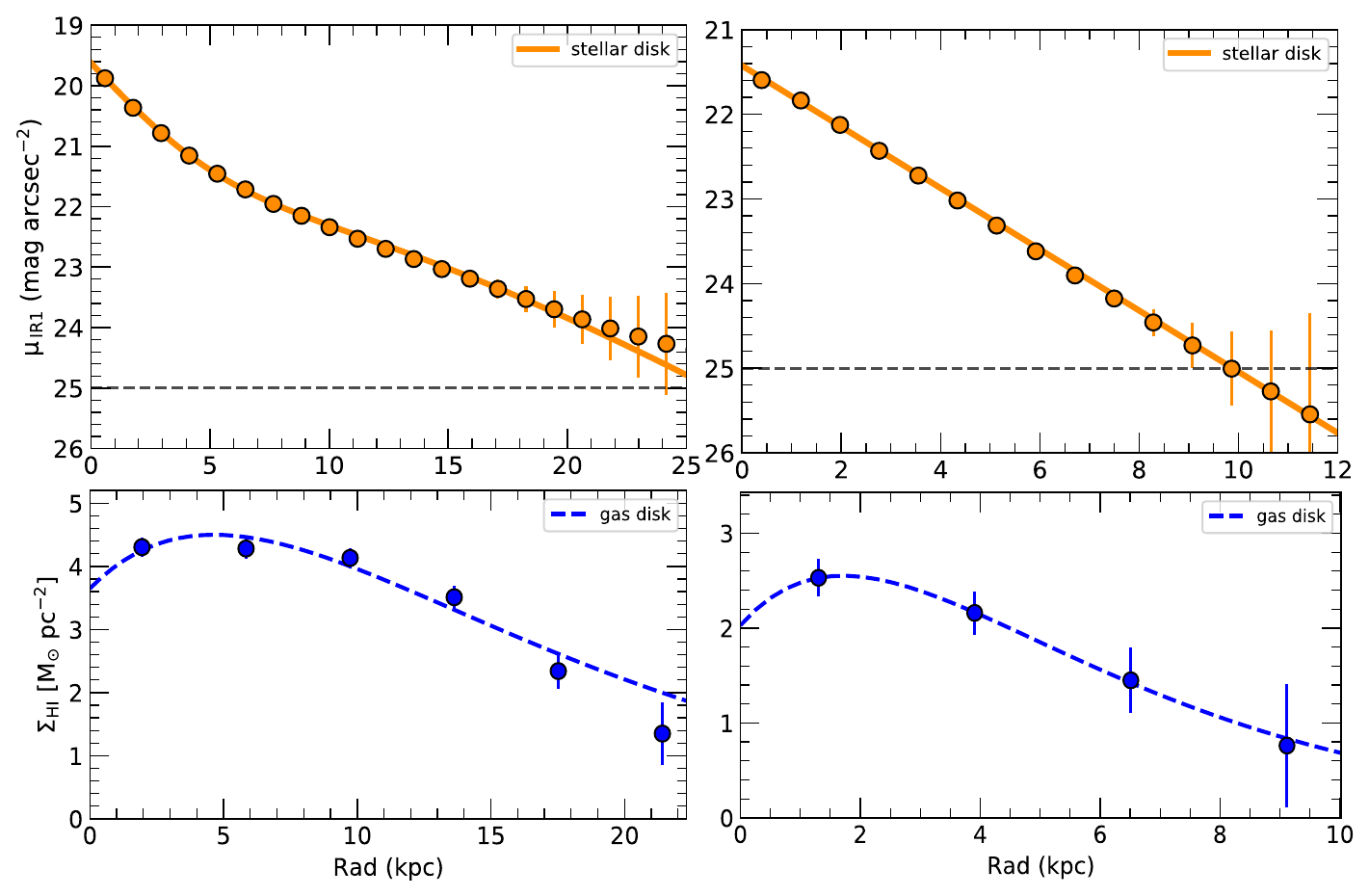}
\caption{Surface brightness and \hi\ gas density profiles for two representative galaxies J095720.6+015507 and J100211.2+020118 from our sample. Left column: Profile for a galaxy with a pseudo bulge (orange points) described by a poly-exponential fit (orange line). The horizontal dashed line represents the 3$\sigma$ magnitude limit. The bottom panel displays the radial profile of the \hi\ gas surface density (blue points), with the dashed blue line representing the poly-exponential gas disc fit and the points indicating the observed data. Right column: Profile for a galaxy without a bulge (orange points), described by exponential disc fit (line), the bottom panel shows the gas surface density profile (blue points) described by the poly-exponential gas disc fit (blue dashed line)}
\label{fig:models_gs}
\end{figure*}

We defined a series of concentric elliptical annuli centred on each galaxy, with the separation between annuli set equal to the Spitzer IRAC 1 point spread function (PSF) full width at half maximum (FWHM $\sim 1.7''$). The number of elliptical apertures was left unconstrained and extended to the edge of the image, where the background is well sampled, ensuring that the integrated light converges and the radial surface-brightness profile reaches a flat asymptotic tail. The ellipticity and position angle of the ellipses were fixed to the geometry measured using Eq.~\ref{inclination}. Fixing the ellipse geometry, rather than allowing each annulus to follow the isophotal contours, has been shown to be effective for deriving surface-brightness profiles, particularly for 3.6 $\mu$m data \citep{marasco2023, sileg2024}.

Within each annulus, the mean intensity was calculated after correcting for residual background flux and applying aperture corrections. The measurement errors on the mean intensity were computed following the standard photometric error-propagation approach, in which the total variance is the sum of contributions from the object signal, sky background, and readout noise, with background noise dominating the uncertainty at faint levels \citep{howell2006}. The resulting surface-brightness profiles, corrected for inclination, are shown in Figure~\ref{fig:sbp} for all galaxies in the sample.

\subsection{Resolved stellar mass surface densities}
 Owing to the limited wavelength coverage of many \hi samples, 3.6~$\mu$m imaging has commonly been used as the optimal single-band proxy for stellar mass. However, within the MIGHTEE fields we have access to some of the deepest and high-quality visible \citep[e.g.][]{Aihara2018, Aihara2019} and near-infrared \citep[e.g.][]{McCracken2012, jarvis2013} imaging data available. 

Therefore, in addition to the stellar surface-brightness profiles derived from the Spitzer IRAC 1 (3.6 $\mu$m) imaging, we incorporate into our analysis the resolved stellar mass surface densities measured by \citet{Varasteanu2025}. These were obtained through spatially resolved spectral energy distribution (SED) fitting across ten optical and near-infrared bands, yielding resolved stellar mass surface-density profiles (see Figure 4 of \citealt{Varasteanu2025}). 

The \citet{Varasteanu2025} sample is composed of the same \hi-selected galaxies and overlaps with our own sample for 19 out of 20 systems. Their resolved stellar mass measurements therefore provide an independent and complementary dataset that strengthens our characterisation of the stellar mass distribution and its contribution to the baryonic mass budget. Spatially resolved stellar mass surface densities based on radially varying $\Upsilon_{*}$ have so far been used for mass modelling and DM inference only in a single-object study \citep{tamm2012}. Here we extend this approach to a larger sample.
Hereafter, we refer to these resolved stellar mass surface-density profiles as the resolved $\Upsilon_{*}$ profiles.

\section{Mass Models}
\label{sec:mass_models}
Rotation curve decomposition enables us to separate the contributions from visible and dark matter, offering a comprehensive picture of the overall mass structure within a galaxy. The total rotational velocity $V_{\mathrm{tot}}(R)$ at a given radius $R$ can be expressed as the square root of the quadratic sum of the rotational velocities contributed by the various mass components of the galaxy. These components include the baryonic matter comprising gas, bulge and disc ($V_{\mathrm{bar}}$), and the dark matter halo ($V_{\mathrm{halo}}$):
\begin{figure*} 
\centering
\includegraphics[scale=0.7]{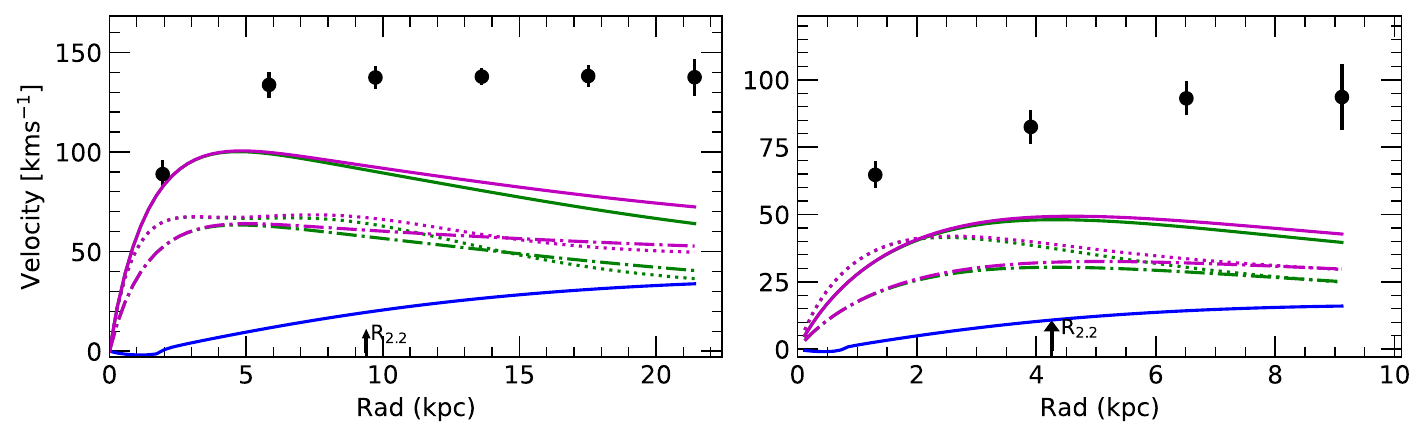}
\caption{Baryonic mass modelling for two representative galaxies J095720.6+015507 and J100211.2+020118 from our sample. Left panel: Mass modelling for the galaxy with poly-exponential disc. The observed rotation curve (black points) is compared with modelled rotational velocities from gas (blue line) and stellar disc (green line). The total baryonic rotation curve $V_{\mathrm{bar}}$ is shown with a magenta line. Solid lines represent adopted $\Upsilon_{*}=0.5$, dashed-dotted lines correspond to $\Upsilon_{*}=0.2$ and dotted lines correspond to the SED profiles. The arrow indicates the radius corresponding to 2.2 disc scale lengths. Right panel: Baryonic mass modelling for the galaxy with pure exponential disc}
\label{fig_mm}
\end{figure*}
\begin{equation}
V^2_{\mathrm{tot}}(R) = V^2_{\mathrm{bar}}(R) + V^2_{\mathrm{halo}}(R).
\end{equation}

Analysing the individual contributions of these components allows us to understand the distribution of mass within each one. The gas component $V_{\mathrm{gas}}(R)$ is derived from the observed \hi\ surface density profile (Section \ref{sec:gas_sbp}), while the stellar components $V_{\mathrm{bulge}}(R)$ and $V_{\mathrm{disc}}(R)$ are obtained from the surface brightness profiles (Section \ref{sec:stellarmass}) converted to mass profiles using appropriate mass-to-light ratios. We neglect the contribution from molecular gas, which is expected to have only a minor impact on the mass modelling\citep{frank2016, ponomareva2018, mancerapina2022b}. The DM halo's contribution $V_{\mathrm{halo}}(R)$ is modelled to account for any discrepancy between the observed rotation curve (Section \ref{sec:RCs}) and the one from the baryonic components:
\begin{equation}
V^2_{\mathrm{bar}}(R) = V^2_{\mathrm{gas}}(R) + V^2_{\mathrm{stars}}(R).
\end{equation}

We construct our mass models by closely following the methodology from \citet{mancerapina2022a, mancerapina2022b}.  To compute the gas and stellar rotation velocities, we utilise the software \textsc{galpynamics}\footnote{\url{https://gitlab.com/iogiul/galpynamics}}, which numerically calculates the gravitational potential of a given mass distribution described by a density profile as a function of radius and height above the disc mid-plane. Subsequently, \textsc{galpynamics} computes the circular velocity of the mass distribution by taking the derivative of the gravitational potential evaluated at the mid-plane of the mass component (see Section 3.1 in \citealt{mancerapina2022b} for details).

To obtain the functional forms of the density distribution of the baryonic components of our sample galaxies, we first fit the observed \hi\ surface mass density profiles with the poly-exponential function:
\begin{equation} 
\Sigma_{\mathrm{HI}}(R) = \Sigma_0 \exp\left(-\frac{R}{R_d}\right) \left(1 + c_1 R + c_n R^{n} + \dots \right), 
\label{eq:polyex}
\end{equation}
where $\Sigma_0$ is the central surface mass density, $R_d$ is the exponential disc scale length, and $c_1$, $c_2$, $c_n$ etc., are the coefficients of the polynomial. 
This function closely matches the surface-density profiles of gas-rich galaxies, which commonly show a central plateau or depression, rise at intermediate radii, and decline approximately exponentially at large radii (e.g. \citealt{swaters1999,wang2016, martinsson2016}). For complex, well-resolved data, the number of polynomial coefficients can reach up to four \citep{pavel2024}. However, for our marginally resolved \hi\ surface mass density profiles, we use only the first-degree coefficient to avoid over-fitting. The resulting fits are shown in Figure~\ref{fig:models_gs} (bottom panels) for two representative galaxies from our sample and demonstrate that the chosen functional form can reproduce the data, capturing the central depression common in the \hi\ surface mass density distribution \citep{ponomareva2016}.
\begin{figure*} 
\centering
\includegraphics[scale=0.7]{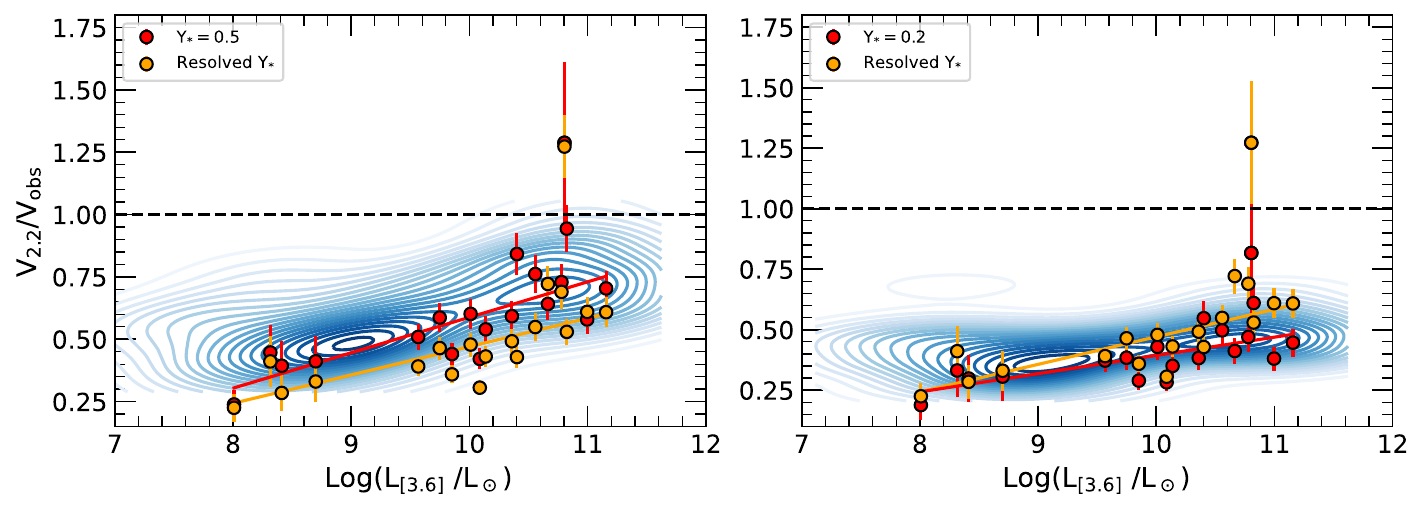}
\caption{Ratio of the baryonic to observed rotation velocity at 2.2 disc scale lengths (V$_{2.2}$/V$_\text{obs}$) as a function of the total luminosity in the 3.6 $\mu$m band for our sample galaxies (red points), shown for two adopted values of the stellar mass-to-light ratio $\Upsilon_{*}$ (Section \ref{sec:mass_models}). Observational trends from the SPARC database \citep{lelli2016} are highlighted with blue regions. The dashed line represents V$_\text{bar}$/V$_{\text obs}=1$, which marks the upper limit for physically acceptable discs
}
\label{fig_mass_model}
\end{figure*}

Next, we model the stellar light distribution by fitting the stellar surface brightness profiles (Figure \ref{fig:sbp}) using a functional form. For galaxies with a deviation from the pure exponential disc and excess light in the centre we first model the total intensity as the sum of two components: a bulge, described by a general S\'ersic profile: 
\begin{equation}
I(R) = I_e \exp\left(-b_n \left[\left(\frac{R}{R_e}\right)^{1/n} - 1\right]\right),
\end{equation}
where $I_e$ is the intensity at the effective radius $R_e$, $n$ is the S\'ersic index, and 
$b_{n}$ is a constant that depends on $n$ \citep{ciotti1999},
and a disc, represented by an exponential profile (a S\'ersic profile with $n=1$). We find that none of the bulges in our sample actually represent the classical bulge population, which is typically described with Sérsic index $n > 2$. Instead, we find that bulges in our sample are consistent with pseudobulges ($n \leq 2$). Pseudobulges, also known as disc-like bulges, are thought to form through the secular evolution of the galaxy's disc, driven by internal processes such as bars, spiral arms, and oval distortions \citep{kormendy2004}. Therefore, we cannot model the dynamics of these components as purely spherical, like classical bulges. Instead, we fit our stellar profiles that exhibit this feature with a poly-exponential disc of the fourth order (Eq.~\ref{eq:polyex}; \citealt{pavel2024}). For pure disc systems, we fit only an exponential disc profile. Figure~\ref{fig:models_gs} (upper panels) shows examples of these fits for two representative galaxies: one with a pseudobulge and another with a purely exponential disc.

To convert Spitzer intensities to stellar mass surface density, we use the prescription from \citet{lelli2016}. We apply two different values of the $\Upsilon_{*}$ for the stellar disc: the traditional value for the 3.6 $\mu$m band, $\Upsilon_{*}=0.5$ \citep{meidt14, schombert2019} and $\Upsilon_{*}=0.2$, found by \citet{martinsson13} for the discMass Survey and by \citet{ponomareva2018} using SED modelling of disc galaxies over 18 photometric bands. In the following, we examine how different choices of $\Upsilon_{*}$ affect the mass modelling and compare these results with those derived from spatially resolved $\Upsilon_{*}$ estimates. 

 We adopt the same functional forms (poly- and exponential disk) to describe the resolved $\Upsilon_{*}$ profiles from \citet{Varasteanu2025}, who demonstrated that this form provides an adequate representation of the data.

To account for the stellar disc thicknesses, we adopt the standard assumption that the scale height of the stellar disk is $\sim 0.1 R_d$ \citep{yoachim2006, comeron2018}. We model the vertical structure of the stellar disc using a $sech^{2}$ density profile
 \citep{vdkruit2011}. For the gas disc, we assume the razor-thin approximation \citep{bosma1981, wang2016}. While in reality the gas discs flare with radius (e.g. \citealt{romeo1992,Bacchini2019}), the dynamical effect of the flaring is only important for gas-rich dwarf galaxies with $V_{\rm rot} \lesssim 30\,\rm{km/s}$ and $V_{\rm rot}/\sigma_v\lesssim3$ \citep{mancerapina2022b,mancerapina2025}, and therefore can be neglected in our sample. Moreover, accurate computation of scale heights requires highly resolved \hi\ observations, which are not yet attainable for samples beyond the very local universe. Therefore, we consider our approximations suitable for our sample and proceed with modelling the dynamics of the baryonic components.

 Figure~\ref{fig_mm} illustrates the baryons-only mass models for two representative galaxies from our sample, with stellar circular velocities computed from the 3.6 $\mu$m profiles assuming $\Upsilon_{*}=0.5$ (solid lines) and $\Upsilon_{*}=0.2$ (dash–dotted lines), and from the resolved $\Upsilon_{*}$ profiles (dotted lines). From Figure~\ref{fig_mm}, it is clear that for the galaxy with higher rotational velocity, the stellar circular velocity derived from the  resolved $\Upsilon_{*}$ is closer to that obtained assuming $\Upsilon_{*}=0.2$, whereas for the galaxy with lower rotational velocity it is closer to the case with $\Upsilon_{*}=0.5$. This behaviour is consistent with the findings of \citet{Varasteanu2025}, who showed that galaxies do not share a universal stellar mass-to-light ratio, and that $\Upsilon_{*}$ varies not only from galaxy to galaxy but also within individual galaxies as a function of radius (see Figures~6 and~7 in \citealt{Varasteanu2025}).

An arrow indicates the radius corresponding to 2.2$\times R_{d}$ (R$_{2.2}$) \citep{freeman1970}. This radius is often used for the mass modelling of galaxies since for an exponential disc, the stellar rotation curve reaches its maximum at $\sim R_{2.2}$ \citep{vanalbada1985, binney2008}. The ratio between the rotational velocity of the baryons and the observed rotational velocity at this characteristic radius gives a measure of the degree of the disc ``maximality'' \citep{martinsson13, starkman2018}. A ``maximal disk'' in a spiral galaxy is commonly defined as one in which the stellar disk component contributes the largest fraction of the observed rotation curve ($\sim 85\%\pm10\%$ of the peak velocity), while leaving minimal room for a DM halo contribution \citep{vanalbada1986}.
Figure~\ref{fig_mm} shows that for the first galaxy (left), $\Upsilon_* = 0.5$ produces a maximal disc, as any higher $\Upsilon_*$ would result in a total rotation curve exceeding the observed data. In contrast, the second galaxy (right) demonstrates greater flexibility, allowing for either a larger $\Upsilon_*$ or a more substantial DM contribution. Adopting $\Upsilon_* = 0.2$, on the other hand, results in the discs of both galaxies being submaximal.

\subsection{The Effect of the Stellar Mass-To-Light Ratio}
Figure \ref{fig_mass_model} shows the ratio of the baryonic to observed rotational velocities ($V_{2.2}/V_{\mathrm{obs}}$) measured at $2.2$R$_d$ for our sample galaxies, considering three different $\Upsilon_{*}$ as described in Section \ref{sec:mass_models}. 

 Firstly, we note that our sample reproduces the same trends found by \citet{lelli2016} for 175 large nearby disc galaxies (the SPARC sample). We find that the degree of disc maximality depends on the total luminosity of a galaxy when $\Upsilon_{*} = 0.5$ is adopted. Figure~\ref{fig_mass_model} (left) demonstrates that the values of $V_{2.2}/V_{\mathrm{obs}}$ increase from approximately 0.25 for faint galaxies to nearly 1 for galaxies with higher total luminosity.

Conversely, this trend almost disappears when $\Upsilon_{*} = 0.2$ is adopted (Figure~\ref{fig_mass_model}, right), signifying little dynamical distinction between galaxies of different luminosities when a low $\Upsilon_{*}$ is assumed. However, a scenario in which all spiral galaxies have uniformly low stellar mass contributions would lead to systematic deviations from the baryonic Tully--Fisher relation \citep{mcgaugh00} at fixed rotation velocity, which are not observed \citep{ponomareva2018, ponomareva2021}.

Moreover, as demonstrated by \citet{Varasteanu2025}, spatially resolved stellar mass surface densities derived from SED fitting reveal that $\Upsilon_{*}$ varies both from galaxy to galaxy and as a function of radius within individual systems. This implies that the stellar contribution to the baryonic mass distribution is neither globally uniform nor radially constant, but instead depends on the specific properties of each galaxy and location within it.

To assess how such non-universal, resolved $\Upsilon_{*}$ affects trends in fractional baryonic contribution in the inner regions, we examine their impact on the $V_{2.2}/V_{\mathrm{obs}}$--luminosity relation. For this purpose, we fit a linear relation in order to quantify how the slopes of the relations change. We find that when a high stellar mass-to-light ratio ($\Upsilon_{\star}=0.5$) is assumed, the ratio $V_{2.2}/V_{\mathrm{obs}}$ increases systematically with luminosity, with a slope of $0.18 \pm 0.04$, indicating that more luminous spiral galaxies are increasingly baryon-dominated in their inner regions. Resolved stellar mass surface densities yield a similarly strong luminosity dependence, with a slope of $0.15 \pm 0.03$, differing primarily in normalisation (Figure \ref{fig_mass_model}, orange). This demonstrates that incorporating spatially resolved stellar mass distributions preserves the physical trend of increasing inner baryonic dominance with luminosity while refining the stellar mass contribution on a galaxy-by-galaxy basis.
In contrast, adopting a low stellar mass-to-light ratio ($\Upsilon_{\star}=0.2$) tends to reduce the baryonic contribution at $2.2R_{\mathrm{d}}$ across the full luminosity range, leading to an almost flat relation with a slope of only $0.05 \pm 0.03$.

Therefore, the resolved stellar mass estimates preserve a physically meaningful $V_{2.2}/V_{\mathrm{obs}}$ dependence on luminosity, as well as alleviate the extreme form of the disc--halo degeneracy introduced by assuming a single $\Upsilon_{\star}$.

\begin{figure*} 
\centering
\includegraphics[scale=0.55]{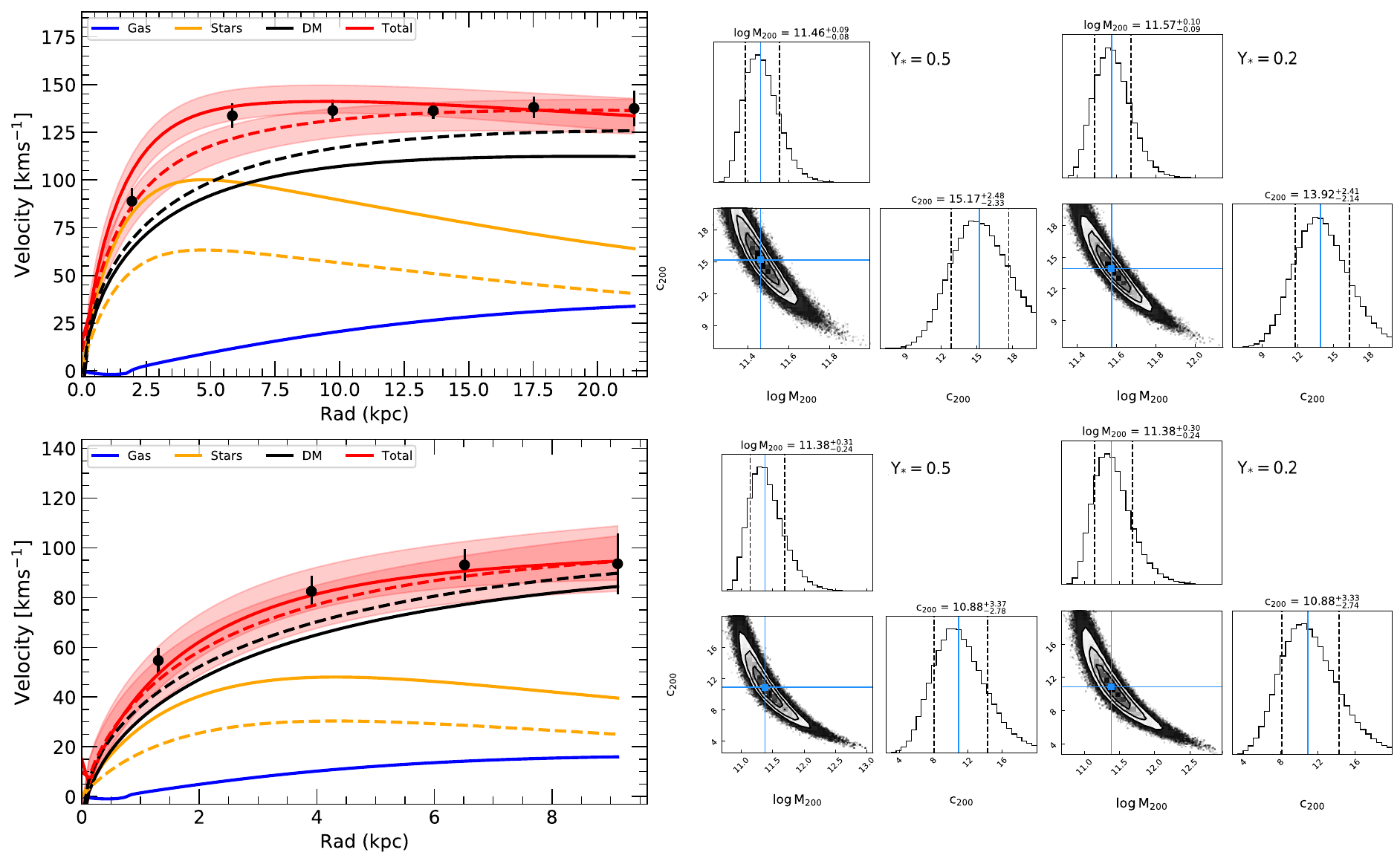}
\caption{Mass modelling of the NFW halo profile for two galaxies J095720.6+015507 and J100211.2+020118. The upper panel shows the rotation curve decomposition for a galaxy with a poly-exponential stellar disc, and the lower panel for a galaxy with a pure exponential disc. The decomposition is shown for fixed $\Upsilon_{*}$ values, with solid lines corresponding to $\Upsilon_{*}=0.5$ and dashed lines to $\Upsilon_{*}=0.2$. The posteriors for these fits are shown on the right.
}
\label{fig_DMG05}
\end{figure*}

\begin{figure*} 
\centering
\includegraphics[scale=0.55]{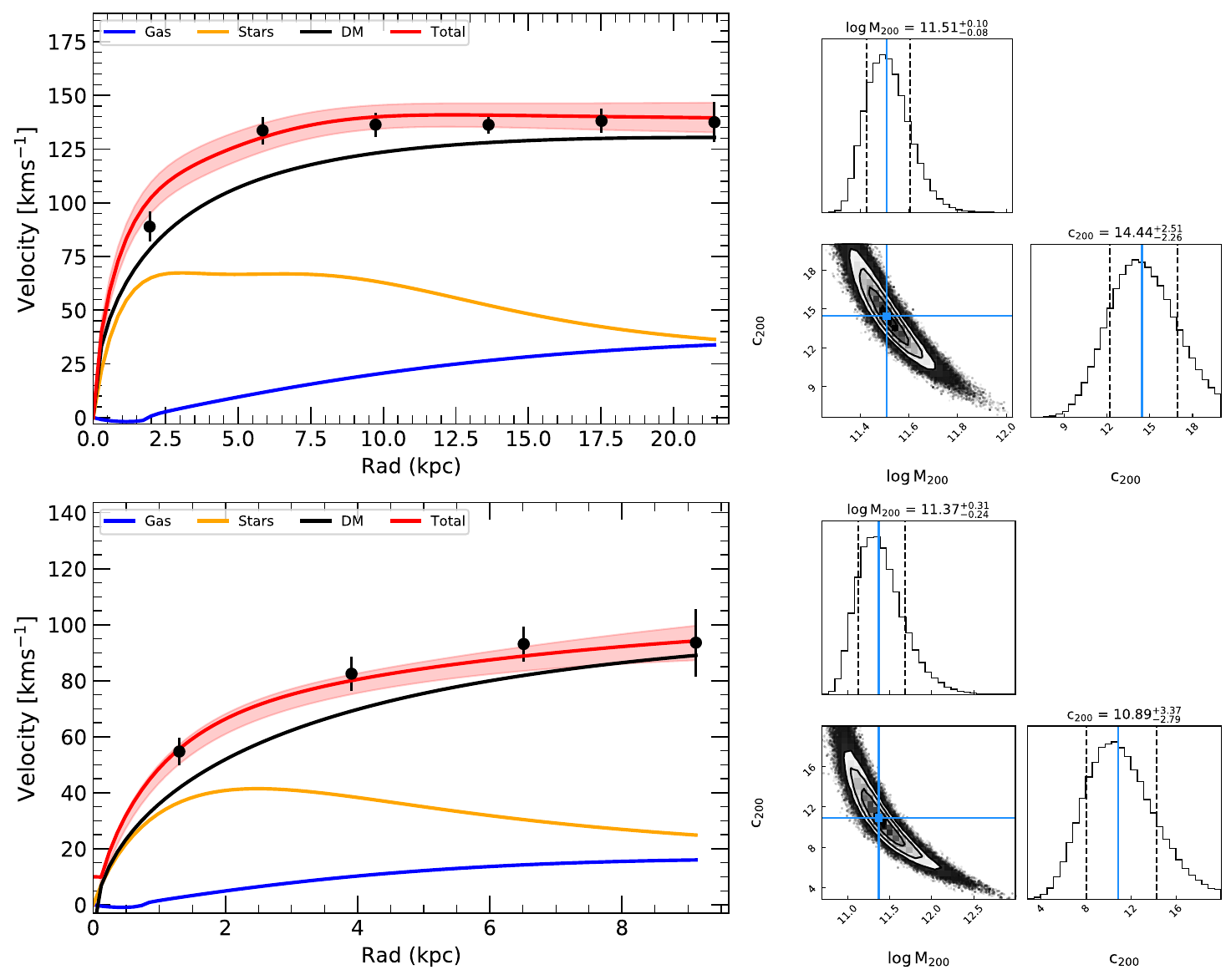}
\caption{Mass modelling of the NFW halo profile for two galaxies J095720.6+015507 and J100211.2+020118 using the resolved stellar surface mass density. The upper panel shows the rotation curve decomposition for a galaxy with a poly-exponential stellar disc, and the lower panel for a galaxy with a pure exponential disc. The posteriors for these fits are shown on the right.
}
\label{fig_DMGres}
\end{figure*}

\section{Dark Matter Properties}
\label{sec:dark_matter}

Given the limited resolution of the \hi rotation curves for our sample, particularly in the inner regions of the galaxies, we only focus on the standard Navarro-Frenk-White (NFW; \citealt{nfw1996}) dark matter halo density profile. Therefore, we do not attempt to differentiate between various DM halo models and instead aim to robustly constrain the DM halo mass and assess the position of our sample galaxies with respect to established DM scaling relations. By restricting the complexity of the halo model, we ensure that the inferred parameters are physically motivated.
Assuming spherical symmetry, the NFW density profile is defined as:
\begin{equation}
\rho = \frac{4\,\rho_s}{\left(\frac{r}{r_s}\right)\left(1+\frac{r}{r_s}\right)^2},
\end{equation}
where $r_s$ is a scale radius, and $\rho_s$ is the characteristic density at $r_s$.
The characteristic density $\rho_s$ is related to the halo mass $M_{200}$ and 
concentration $c_{200}$ as
\begin{equation}
\rho_s = \frac{M_{200}}{16\pi\,r_s^3\left[\ln(1 + c_{200}) - \frac{c_{200}}{1 + c_{200}}\right]},
\end{equation}
where $M_{200}$ is the mass enclosed within radius $R_{200}$, defined such that the 
average density inside $R_{200}$ is $200\,\rho_{\rm crit}$ (with $\rho_{\rm crit}$ the 
critical density of the Universe), and $c_{200} \equiv R_{200}/r_s$ is the concentration parameter.
The enclosed mass at radius $r$ for an NFW halo can then be expressed through the dimensionless 
variable $x = r/r_s$:
\begin{equation}
M(r) = M_{200}\frac{\ln(1+x)-\frac{x}{1+x}}{\ln(1+c_{200})-\frac{c_{200}}{1+c_{200}}}.
\end{equation}
Following the notion of spherical symmetry, the circular velocity at radius $r$ is defined by $V_{\rm NFW}(r) = \sqrt{G M(r)/r}$. 
Introducing the characteristic velocity $V_{200} = \sqrt{G M_{200}/R_{200}}$, we 
can rewrite $V_{\rm NFW}(r)$ in terms of $c_{200}$ and $x$:
\begin{equation}
V_{\rm NFW}(r) = V_{200}\,\sqrt{\frac{c_{200}}{x}}\frac{\sqrt{\ln(1+x)-\frac{x}{1+x}}}{\sqrt{\ln(1+c_{200})-\frac{c_{200}}{1+c_{200}}}}.
\end{equation}
The total (baryons and DM) circular velocity of a galaxy at each radius is then:
\begin{equation}
V_{\rm tot}^2(R) = V_{\rm NFW}^2(R) + V^2_{\rm disc}(R) +V^2_{\mathrm{gas}}(R),
\end{equation}

\begin{figure*} 
\centering
\includegraphics[scale=0.6]{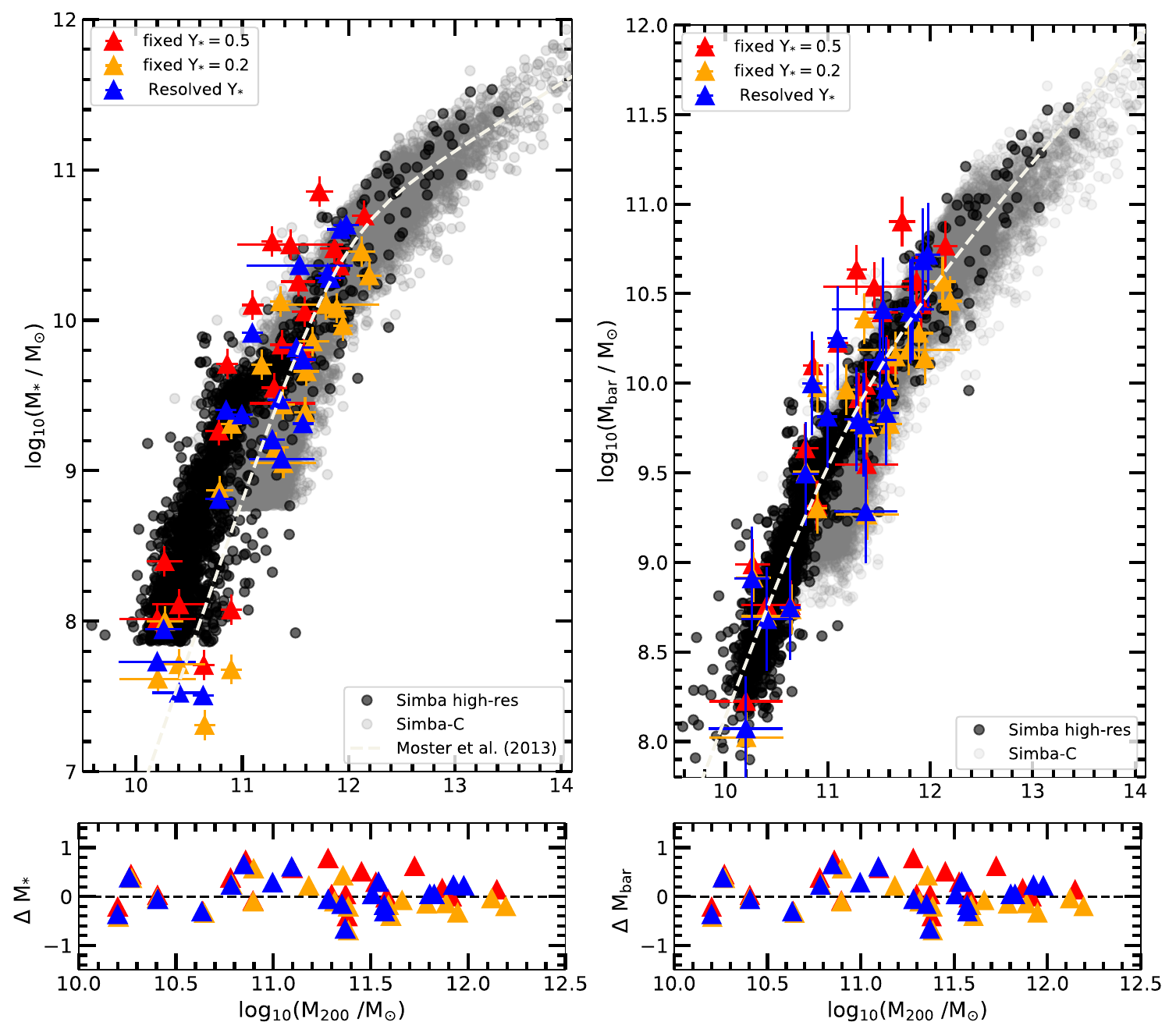}
\caption{Left: Stellar-to-halo mass relation (SHMR) for our sample galaxies derived under different $\Upsilon_{*}$ assumptions. Red triangles indicate results obtained with fixed $\Upsilon_{*}=0.5$, orange symbols correspond to fixed $\Upsilon_{*}=0.2$, and blue symbols represent models where $\Upsilon_{*}$ is left free. Predictions from the SIMBA-C cosmological hydrodynamical simulations are shown as grey points, while the SIMBA high-resolution run is represented by black points. The abundance-matching relation from \citet{moster2013} is shown as the dashed white line.
Right: Baryonic-to-halo mass relation (BHMR) for the same sample. The dashed white line indicates the fitted relation to the theoretical models. The lower panels show the residuals with respect to the corresponding theoretical relations. }
\label{fig_M200}
\end{figure*}

To calculate $V_{\rm tot}$ we adopt a standard Bayesian mass-modelling framework combined with Markov Chain Monte Carlo (MCMC) sampling \citep{posti2019, mancerapina2022b, diteodoro2023}, using the Python implementation $emcee$ \citep{emcee}. By employing an MCMC sampler, we explore the parameter space defined by halo mass ($M_{200}$) and halo concentration ($c_{200}$). Our goal is to constrain these key parameters by fitting theoretical rotation curves to the observed galaxy kinematics. We assume a $\Lambda$CDM cosmology and adopt standard values for the critical density ($\rho_{\rm crit}$) and gravitational constant ($G$). We use a standard $\chi^2$ likelihood function $\mathcal{L}$, defined as:
\begin{equation}
\chi^2 = -2\ln \mathcal{L}(V_{\rm rot} \mid V_{\rm mod}(x)) = \sum_{i=1}^{N} \left(\frac{V_{\rm rot}(r_i) - V_{\rm mod}(r_i)}{\Delta_i}\right)^2,
\end{equation}
where $V_{\rm rot}(r_i)$ and $V_{\rm mod}(r_i)$ are the observed and modelled circular velocities at the $i$-th radius $r_i$, respectively, and $\Delta_i$ are the uncertainties on the rotation velocity \citep{diteodoro2023}.

We impose physically motivated priors on the fitted parameters during the MCMC sampling. For $M_{200}$ we use a uniform prior over a wide range $6 < \log\, (M_{\rm h}/M_{\odot}) \leq 15.$ For the halo concentration instead, we use a lognormal prior based on the mass-concentration relation derived from simulations, enforcing that $\log_{10}(c_{200})$ follows a normal distribution with a mass-dependent mean and a fixed standard deviation of $0.11$ dex \citep{dutton14}.  
This choice is motivated by the studies of \citet{katz2017,posti2019}, who have shown that even with high-quality local data a strong non-uniform prior on $c_{200}$ is necessary to obtain proper constraints on the DM halo parameters (but see also \citealt{mancerapina2022b}). Indeed, in Cold Dark Matter (CDM) cosmological simulations, the halo concentration $c_{200}$ is not independent of $M_{200}$. Instead, there is a well-known mass-concentration relation, typically showing that higher-mass halos tend to have lower concentrations. This anti-correlation arises from the formation history of halos. Less massive halos typically collapse earlier, when the mean background density is higher. Early collapse leads to denser central regions and thus higher concentrations. More massive halos, on the other hand, often form later through mergers and the accretion of matter from a lower-density environment. As a result, their central regions are less compact, resulting in lower concentrations. This relation encodes information about the formation epoch and assembly history of halos, showing that their structure cannot be understood as simple scaled versions of one another \citep{ludlow14, dutton14, diemer19}.

For each fitting run we used 100 walkers for 5000 steps, discarding the first 20\% as a burn-in phase. We verify that our fits have converged by computing the integrated autocorrelation time $\tau$ and ensuring that each walker contributes at least 50$\tau$ samples. Each fitted parameter’s central value was taken as the median of its posterior (50th percentile), and uncertainties were defined by the 16th and 84th percentiles. 
We perform three modelling runs for each galaxy, one with resolved $\Upsilon_{*}$, and another two with $\Upsilon_{*}$ fixed at values of 0.2 and 0.5, as described in Section \ref{sec:mass_models}. 

To assess the quality of the fit and model complexity, we use the Bayesian Information Criterion (BIC):
\begin{equation}
    \text{BIC} = -2 \ln(\hat{\mathcal{L}}) + k \ln(n)
\end{equation}
where $\hat{\mathcal{L}}$ is the maximum value of the likelihood function for the model, $k$ is the number of free parameters and $n$ is the number of data points. The BIC provides a means to compare different models and assumptions about the most suitable $\Upsilon_{*}$. 

 The results for two representative galaxies using $\Upsilon_{*}=0.5$ and $\Upsilon_{*}=0.2$ are shown in Figure~\ref{fig_DMG05}. The figure presents the rotation-curve decomposition, alongside the posterior distributions of the fitted parameters. The $M_{200}$ and $c_{200}$ parameters are tightly constrained, with the posterior distribution of the latter reflecting the imposed mass--concentration relation. This indicates that the dark-matter halo properties are well constrained by the data, largely independently of the adopted stellar mass-to-light ratio. The BIC generally favours the models with $\Upsilon_{*}=0.5$, with $\langle \mathrm{BIC}_{0.5}-\mathrm{BIC}_{0.2} \rangle = -29$ \citep{Kass}. From Figure~\ref{fig_DMG05}, it is clear that the main difference between the total rotation curves obtained using the two different $\Upsilon_{*}$ values is confined to the central regions of the galaxies. At larger radii, the models converge and therefore reproduce the outer rotation velocities equally well, demonstrating that uncertainties in the stellar mass prescription primarily affect the inner mass distribution, while the outer rotation curve is dominated by the halo.

Figure~\ref{fig_DMGres} shows the same rotation-curve decomposition for the same galaxies, but using resolved stellar surface mass density profiles rather than the IRAC~1 surface-brightness profiles scaled by a fixed $\Upsilon_{*}$. 
Overall, model comparison based on the BIC shows a strong preference for the resolved-$\Upsilon_{*}$ prescription, with mean offsets of $\langle \mathrm{BIC}{\rm resolved}-\mathrm{BIC}{0.2} \rangle = -47$ and $\langle \mathrm{BIC}{\rm resolved}-\mathrm{BIC}{0.5} \rangle = -15$. 
From Figure~\ref{fig_DMGres} (upper panel), we can see that the model reproduces the observed rotation curve and its features more accurately when the resolved stellar mass surface density is used, with $\mathrm{BIC}_{\rm resolved} = 597$, compared to $\mathrm{BIC}_{0.5} = 658$ and $\mathrm{BIC}_{0.2} = 834$. This improvement reflects the ability of the resolved stellar mass distribution to capture radial variations in the stellar contribution, rather than simply rescaling a fixed light profile, and demonstrates that incorporating spatially resolved stellar mass information leads to a more accurate description of the observed rotation curve, although we note that the best-fit parameters for $M_{200}$ and $c_{200}$ are formally consistent across all tested values of $\Upsilon_{*}$. The $M_{200}$ obtained using different $\Upsilon_{*}$ prescriptions are listed in Table \ref{tbl_MM}.

\subsection{Halo mass scaling relations}
Having derived the halo masses from our rotation curve decompositions, we now examine how these halos relate to their stellar and baryonic content. By combining our halo mass estimates with stellar masses, we first explore the stellar-to-halo mass relation (SHMR). The SHMR describes how the stellar mass of galaxies is connected to the mass of their surrounding DM halo. In essence, it quantifies the efficiency with which baryonic matter is converted into stars at different halo mass scales. Lower-mass halos tend to be less efficient at forming stars (e.g. \citealt{posti2019,mancerapina2025}), presumably due to a combination of factors such as supernova feedback and reionization, while intermediate-mass halos are often thought to be most efficient, and very massive halos may again show reduced star formation efficiency, potentially due to active galactic nuclei feedback and virial shock heating \citep{vale2008,conroy2009,moster10}. In theoretical work, the SHMR is usually derived via abundance matching (AM) between stellar mass functions and dark matter-only simulations \citep{moster2013,behroozi2013,behroozi2019}, while observational analyses test whether these relations hold for real galaxies \citep{posti2019, diteodoro2023}. These studies highlight the importance of feedback processes and environmental effects in shaping the interplay between dark matter and baryons. The SHMR, therefore, links the growth of stellar mass in galaxies to the assembly of their dark matter halos, providing critical tests for theoretical models of galaxy formation and evolution. 

Figure \ref{fig_M200} (left) illustrates the SHMR for our sample, computed under three different assumptions for the stellar mass. We quantify the impact of different stellar-mass prescriptions on the inferred halo mass by performing object-by-object comparisons of halo masses obtained from rotation-curve decomposition for the same galaxies, and by characterising the resulting differences through their median offsets and dispersion. Comparing the two fixed mass-to-light ratio prescriptions ($\Upsilon_{*} = 0.5$ and $\Upsilon_{*} = 0.2$) provides a direct measure of the classical disk--halo degeneracy: increasing the stellar mass through a higher global $\Upsilon_{*}$ leads to a systematic, although weak, decrease in the inferred halo mass. In our sample, a change of $\sim$0.4 dex in stellar mass corresponds to a median decrease of $\sim$0.05 dex in $M_{200}$, with this inverse trend present across the galaxy population, demonstrating that while the inner mass decomposition remains degenerate, the total halo mass is only weakly sensitive to even extreme assumptions about the global stellar mass-to-light ratio.

The behaviour changes when adopting spatially resolved stellar-mass distributions. In this case, despite a systematic reduction in stellar mass, the inferred halo masses do not show a coherent decrease relative to the fixed-$\Upsilon_{*}$ models. Instead, the median shift in $M_{200}$ is small ($\sim$0.01 dex), more than an order of magnitude smaller than the intrinsic object-to-object scatter ($\sim$0.08 dex), indicating that the differences are dominated by galaxy-to-galaxy variation rather than a population-level offset. This demonstrates that allowing $\Upsilon_{*}$ to vary with radius does not result in a simple global trade-off between stellar and halo components, but instead introduces galaxy-specific responses that lead to a non-uniform impact on the inferred halo mass.

\begin{table}
\centering
\begin{tabular}{lcc|cc}
\hline
 & \multicolumn{2}{c|}{Med $\Delta$} & \multicolumn{2}{c}{$\sigma$} \\
\cline{2-5}
 & $M_{\star}$ & $M_{bar}$ & $M_{\star}$ & $M_{bar}$ \\
\hline
$\Upsilon_{*}=0.5$ & +0.39  & +0.09  &0.51  & 0.33 \\
$\Upsilon_{*}=0.2$ & -0.15  & -0.12 & 0.48 & 0.30 \\
Resolved $\Upsilon_{*}$ & +0.08  & 0.04 &0.43  & 0.29 \\
\hline
\end{tabular}
\caption{Table summarising the median offsets ($\Delta$) from the theoretical SHMR and BHMR relations for our sample under different $\Upsilon_{*}$ assumptions (Figure~\ref{fig_M200}), along with the observed scatter ($\sigma$) relative to the theoretical relations. All values are expressed in dex.}
\label{tab:model_delta}
\end{table}

To assess which stellar-mass prescription is most consistent with theoretical expectations, we compare the three samples to the stellar–halo mass relation of \citet{moster2013} by evaluating the residuals in $\log M_*$ at fixed $M_{200}$. The fixed $\Upsilon_*=0.5$ sample lies systematically above the relation, with a median offset of $+0.39$ dex, while the fixed $\Upsilon_*=0.2$ sample lies below it, with a median offset of $-0.15$ dex. In contrast, the sample based on spatially resolved stellar-mass distributions is much more closely aligned with the theoretical relation, exhibiting a median offset of only $+0.08$ dex. This offset is smaller in magnitude than for either fixed-$\Upsilon_*$ case and is comparable to the intrinsic scatter of the relation, indicating that the resolved-$\Upsilon_*$ prescription yields stellar–halo mass measurements that are most consistent with the canonical abundance-matching expectation.
For visualisation purposes, we also include data from the hydrodynamic cosmological simulations \textsc{Simba} (\textsc{Simba-C} and high resolution run; \citealt{Dave2019}), which show overall consistency with the abundance-matching relation of \citet{moster2013} across the considered mass range (Figure~\ref{fig_M200}, left).

Incorporating \hi gas mass measurements, we also construct the baryonic-to-halo mass relation (BHMR). BHMR extends the concept of the SHMR by considering not only the stellar mass but also the cold gas mass (M$_*+$1.4\,M$\rm_{HI}$) to form a more complete picture of a galaxy’s total baryon content relative to its host halo. While the SHMR focuses on the efficiency of converting baryons into stars, the baryonic mass-to-halo mass relation captures both the accumulation and retention of baryons, as well as their conversion into stars. By examining the baryonic content, rather than just the stellar component, we can gain insight into how efficient galaxies are at acquiring gas, how much of that gas is converted into stars, and how much is lost through outflows \citep{papastergis2012, mcgaugh2015,mancerapina2025}. 

Figure \ref{fig_M200} (right) presents the BHMR for our sample under the three $\Upsilon_{*}$ assumptions, alongside the BHMR derived from the \simba\ cosmological simulations. From \simba\, we consider two runs: the `flagship' 100~Mpc/$h$ box run (minimum stellar mass resolution of 5.85$\times$10$^{8}$ M$_{\odot}$ for all galaxies considered) from the recently updated \simba-C simulation \citep{simbac2023}, which incorporates improved chemical enrichment models, and a high-resolution \simba\ run designed to resolve low-mass galaxies (minimum stellar mass resolution of 7.30$\times$10$^{7}$ M$_{\odot}$), similar to those in our sample. We limit the comparison to central galaxies, rather than including satellites to better match the MIGHTEE-HI sample.

 As there is no established reference relation for the BHMR, we fit a double power-law relation to the high-resolution \textsc{Simba} sample to provide a baseline against which our observational samples can be compared (Figure \ref{fig_M200}, right). The best-fitting double power-law parameters are 
$\log M_{1} = 11.17 \pm 0.20$, $\log N = -1.44 \pm 0.03$ ($N = 0.0364$), 
$\beta = 0.60 \pm 0.08$, and $\gamma = 0.34 \pm 0.07$, where $M_{1}$ is the 
characteristic halo mass at which the relation transitions between the 
low- and high-mass regimes, $N$ sets the overall normalisation, and 
$\beta$ and $\gamma$ describe the low- and high-mass slopes, respectively. Relative to this relation the sample assuming a fixed $\Upsilon_*=0.5$ lies systematically above the fitted relation, with a median offset of $\sim$0.09 dex, while the fixed $\Upsilon_*=0.2$ sample lies below it, with a median offset of $\sim$-0.12 dex. In contrast, the sample based on spatially resolved stellar-mass estimates shows the closest agreement, exhibiting a median offset of only $\sim$0.04 dex. 
This behaviour is qualitatively similar to that found for the SHMR AM model, where different stellar-mass assumptions systematically shift galaxies above or below the mean relation, while prescriptions that more accurately capture the stellar mass distribution yield the closest correspondence to the theoretical predictions. It is important to note that several studies have reported deviations from the AM predictions, particularly at high mass scales \citep{posti2019, diteodoro2023}, but see \citet{desmond2015,Boreiko2026} for analyses of how sample selection may impact this. We do not observe such departures within the mass range probed by our sample. This further motivates the use of spatially resolved SED-based mass-to-light ratios in upcoming larger samples. The resulting median offsets and scatter of each sample relative to the theoretical relations are listed in Table \ref{tab:model_delta}.

\section{Summary and Conclusions}
\label{sec:summary}
In this paper, we use deep \hi\ observations from the MIGHTEE survey to derive high-quality \hi\ rotation curves for a sample of 20 galaxies in the COSMOS field across a previously unexplored redshift range, $0 \leq z \leq 0.08$. Combining \hi\ rotation curves with carefully derived stellar surface brightness profiles from Spitzer IRAC 1 (3.6~$\mu$m) band data and resolved stellar mass surface densities profiles from \citet{Varasteanu2025}, we explore mass models for our sample galaxies using the rotation curve decomposition technique. We derived DM halo masses by fitting the observed rotation curves with a parametric NFW halo model \citep{nfw1996}, combined with our baryonic mass profiles. Using a Bayesian Markov Chain Monte Carlo (MCMC) approach, we explored the parameter space to find the best-fitting halo mass and concentration. By decomposing the \hi\ rotation curves into their baryonic and dark matter components, we obtain observational constraints on the DM halo masses of our sample galaxies and investigate how the DM halos connect to their stellar and baryonic content.

Our main results can be summarised as follows:
\begin{itemize}
\item Examining the ratio of the baryonic velocity contribution ($V_{\rm bar}$) to the total observed velocity ($V_{\rm obs}$) at the characteristic radius $R_{2.2}$, we find that adopting a higher stellar mass-to-light ratio ($\Upsilon_{*}=0.5$) produces a clear dependence of $V_{2.2}/V_{\rm obs}$ on total galaxy luminosity, consistent with previous studies of nearby disc galaxies \citep{lelli2016}. In this case, more luminous systems approach $V_{\rm bar} \approx V_{\rm obs}$ at $R_{2.2}$, indicating an increased relative baryonic contribution in the inner regions. In contrast, this luminosity dependence largely disappears when a lower value of $\Upsilon_{*}=0.2$ is assumed, resulting in an almost flat $V_{2.2}/V_{\rm obs}$--luminosity relation (Figure~\ref{fig_mass_model}). This demonstrates that the inferred dynamical role of baryons is highly sensitive to the assumed stellar mass-to-light ratio.

Crucially, when resolved stellar mass surface density profiles are used instead of a fixed $\Upsilon_{*}$, the physically meaningful dependence of $V_{2.2}/V_{\rm obs}$ on luminosity is preserved, with a slope comparable to that obtained for $\Upsilon_{*}=0.5$, while avoiding the suppression of this trend seen for $\Upsilon_{*}=0.2$. This shows that allowing for galaxy-dependent and radially varying stellar mass-to-light ratios provides a more realistic description of the stellar contribution to galaxy dynamics. These results highlight that the disc--halo degeneracy is strongly amplified by the assumption of a universal stellar mass-to-light ratio, and that incorporating resolved stellar mass information offers a more physically motivated framework for interpreting rotation curves in the context of galaxy formation models and the interplay between baryonic and dark matter.

\item Comparing rotation-curve decompositions based on fixed and resolved stellar mass prescriptions, we find that differences between models with fixed $\Upsilon_{*}$ are confined to the central regions, while the outer rotation curves are reproduced equally well, indicating that uncertainties in the stellar mass primarily affect the inner mass distribution (Figure \ref{fig_DMG05}). Using resolved stellar surface mass density profiles yields a statistically preferred description of the data, improving the reproduction of detailed rotation-curve features (Figure \ref{fig_DMGres}). This demonstrates that incorporating spatially resolved stellar mass information provides a more accurate and physically motivated representation of galaxy rotation curves than globally scaled light profiles.

\item By comparing stellar--halo mass relations derived under different stellar-mass prescriptions, we show that the inferred total halo mass is only weakly sensitive to even extreme assumptions about a global stellar mass-to-light ratio. While fixed-$\Upsilon_{*}$ models exhibit the expected inverse response between stellar and halo mass, the resulting shifts in $M_{200}$ are small compared to the intrinsic galaxy-to-galaxy scatter. Adopting spatially resolved stellar-mass distributions further alters this behaviour: variations in stellar mass do not translate into a coherent population-level shift in halo mass, but instead produce galaxy-specific responses in the mass decomposition (Figure \ref{fig_M200} left). When compared to the theoretical abundance-matching expectations \citep{moster10}, the resolved-stellar-mass case yields the smallest residuals and the closest overall agreement, indicating that allowing for radial variations in the stellar mass provides a more internally consistent mapping between stellar and halo mass than either fixed-$\Upsilon_{*}$ assumption.

\item By incorporating H\,\textsc{i} gas mass measurements, we extend our analysis to the baryonic--halo mass relation (BHMR), which provides a more complete view of the total baryon content of galaxies relative to their host haloes. Comparing our observational BHMR to \textsc{Simba} simulations, we find that different stellar mass-to-light ratio assumptions systematically shift galaxies relative to the simulation-based reference relation: fixed $\Upsilon_{*}=0.5$ places galaxies above the relation, while $\Upsilon_{*}=0.2$ shifts them below it. In contrast, stellar masses derived from spatially resolved measurements yield the smallest offsets and the closest overall agreement with the fitted \textsc{Simba} baseline (Figure \ref{fig_M200} right). This behaviour mirrors that observed for the SHMR, indicating that prescriptions which more accurately capture the stellar mass distribution also provide the most consistent mapping between baryonic mass and halo mass, and minimise systematic biases when comparing observations to theoretical predictions.

\end{itemize}

In conclusion, studies of galaxy rotation curves and mass modelling provide key insights into the nature of dark matter, the distribution of baryonic matter, and the physical processes governing galaxy formation and evolution. With current and forthcoming observational facilities, such as the SKA, extending these analyses to higher redshifts will enable us to trace the evolution of mass distributions over cosmic time, offering stringent tests of cosmological models.

 In this paper, we have demonstrated that employing spatially resolved spectral energy distribution (SED) fitting across a broad wavelength range provides a powerful avenue for mitigating the disc--halo degeneracy. By mapping stellar populations and dust content across galaxies, resolved SED modelling yields spatially resolved stellar mass surface density profiles, removing the need to assume a single, global mass-to-light ratio. This multi-wavelength approach, spanning ultraviolet to infrared bands, naturally accounts for variations in stellar population age, metallicity, and dust attenuation, all of which have a strong impact on inferred mass-to-light ratios. Incorporating this information into mass modelling leads to a more physically motivated characterisation of the stellar mass distribution and enables a clearer separation between baryonic and dark-matter contributions to galaxy rotation curves.

At the same time, important observational challenges remain. The limited spatial resolution of current \hi\ data beyond the very local universe \citep{ponomareva2021}, together with the signal-to-noise limitations of large \hi\ surveys \citep{deg2024}, restrict our ability to probe the gravitational potential in the innermost regions of galaxies. As a result, distinguishing robustly between different dark-matter halo models, such as cored versus cuspy profiles, remains difficult \citep{deblok97, kurapati2020,mancerapina2025}. Future progress will therefore rely on multi-wavelength synergies, particularly the combination of \hi\ data with high-resolution optical emission-line observations (e.g.\ H$\alpha$) that trace the inner galaxy regions \citep{diteodoro2023}. Such joint analyses will improve constraints on the central mass distribution, further reduce modelling degeneracies, and provide stronger tests of dark-matter halo structure.

\section*{acknowledgements}
We thank the anonymous referee for helpful comments that significantly improved this paper.

AAP is grateful to Giuliano Iorio for making \textsc{galpynamics} publicly available and for providing support in its use.

The MeerKAT telescope is operated by the
South African Radio Astronomy Observatory, which is a facility of
the National Research Foundation, an agency of the Department of
Science and Innovation. We acknowledge use of the Inter-University
Institute for Data Intensive Astronomy (IDIA) data intensive research
cloud for data processing. IDIA is a South African university partnership involving the University of Cape Town, the University of Pretoria and the University of the Western Cape. The authors acknowledge the Centre for High Performance Computing (CHPC), South Africa, for
providing computational resources to this research project. 

This work is based in part on archival data obtained with the Spitzer Space Telescope, which was operated by the Jet Propulsion Laboratory, California Institute of Technology under a contract with NASA. Support for this work was provided by an award issued by JPL/Caltech.

This work is based in part on data products 
produced at Terapix available at the Canadian Astronomy Data
Centre as part of the Canada-France-Hawaii Telescope Legacy Survey,
 a collaborative project of NRC and CNRS. The Hyper SuprimeCam (HSC) 
 collaboration includes the astronomical communities of
Japan and Taiwan, and Princeton University. The HSC instrumentation and 
software were developed by the National Astronomical
Observatory of Japan (NAOJ), the Kavli Institute for the Physics
and Mathematics of the Universe (Kavli IPMU), the University of
Tokyo, the High Energy Accelerator Research Organization (KEK),
the Academia Sinica Institute for Astronomy and Astrophysics in Taiwan 
(ASIAA), and Princeton University. Funding was contributed by
the FIRST program from Japanese Cabinet Office, the Ministry of Education, 
Culture, Sports, Science and Technology (MEXT), the Japan
Society for the Promotion of Science (JSPS), Japan Science and Technology 
Agency (JST), the Toray Science Foundation, NAOJ, Kavli
IPMU, KEK, ASIAA, and Princeton University.

AAP, MJJ and IH acknowledge support of the STFC consolidated grant ST/S000488/1. 
MJJ, IH, AV, TY, HP and SK acknowledge the support of a UKRI Frontiers Research Grant [EP/X026639/1], which was selected by the European Research Council.
MJJ and AAP acknowledge support from the Oxford Hintze Centre for Astrophysical Surveys, which is funded through generous support from the Hintze Family Charitable Foundation. PEMP acknowledges the support from the Dutch Research Council (NWO) through the Veni grant VI.Veni.222.364. IH acknowledge support from the South African Radio Astronomy Observatory (SARAO) which is a facility of the National Research Foundation (NRF), an agency of the Department of Science and Innovation.
MG is supported by the UK STFC Grant ST/Y001117/1. MG acknowledges support from the Inter-University Institute for Data Intensive Astronomy (IDIA). HD is supported by a Royal Society University Research Fellowship (grant no. 211046). KAO acknowledges support by the Royal Society through a Dorothy Hodgkin Fellowship (DHF/R1/231105). IP acknowledges support from the Italian Ministry of Foreign Affairs and International Cooperation (grant number PGR ZA23GR03) and from INAF under the Large Grant 2022 funding scheme (project "MeerKAT and LOFAR Team up: a Unique Radio Window on Galaxy/AGN co-Evolution"). M.B. and A.G. gratefully acknowledge the financial support from the Flemish Fund for Scientific Research (FWO-Vlaanderen) and the South African National Research Foundation (NRF) under their Bilateral Scientific Cooperation program (grant G0G0420N). They also acknowledge the support of networking activities by NRF and the Belgian Science Policy Office (BELSPO), under grant BL/02/SA12 (GALSIMAS).

For the purpose of open access, the author has applied a Creative Commons Attribution (CC BY) licence to any Author Accepted Manuscript version arising from this submission.

This research has made use of NASA’s Astrophysics Data System Bibliographic Services. This research made use of Astropy\footnote{\href{http://www.astropy.org}{http://www.astropy.org}}, a community-developed core Python package for Astronomy.

\section*{Data availability}
The raw visibility data are publicly available from the SARAO archive by searching for the capture block IDs listed in Table 1 of \cite{heywood2024}. The Spitzer IRAC 1 images used in this work are also publicly accessible as part of S-COSMOS \citep{sanders2007}. Tables summarising the main properties of the sample and the derived parameters are provided in the Appendix (Tables \ref{tbl_sample} and \ref{tbl_MM}). All data involved in the analysis are available upon reasonable request to the corresponding author.
\bibliographystyle{mnras}
\bibstyle{mnras}
\bibliography{MIGHTEE_MM}
\vspace*{1cm}

\appendix
\section{Sample Data Tables}
Tables summarising the main properties of our sample \ref{tbl_sample} and the results of the mass modelling \ref{tbl_MM}


\begin{table*}
\centering
\begin{adjustbox}{width=1\textwidth}
\begin{tabular}{lcccccccccc}
\hline
\hline
Name& ra&dec&z&log$_{10}$(MHI)&$i^{\circ}$& h$_d$& pa & Nb&log$_{10}$(L$_{\rm IR1}$)&log$_{10}$(M$_{\star})$\\	
   & deg & deg & & M$_{\odot}$ & deg& kpc& deg& &L$_{\odot}$ & M$_{\odot}$ \\
 \hline
J095829.1+014139&149.621&1.694&0.006&$9.144\pm0.008$&66.759&$1.933\pm0.251$&60.752   & 11&$8.378\pm0.137$&	  --           \Tstrut\\
J095846.8+022051&149.695&2.348&0.006&$8.575\pm0.015$&69.235&$1.034\pm0.134$&   311.745  & 7&$8.008\pm0.133$&	$7.485  \pm 0.088$ \Tstrut\\
J095927.9+020025&149.866&2.007&0.013&$8.505\pm0.065$&40.824&$0.972\pm0.126$&  293.299   & 3&$8.412\pm0.137$&	$7.653  \pm 0.101$ \Tstrut\\
J100005.8+015440&150.024&1.911&0.006&$7.661\pm0.126$&65.857&$0.794\pm0.103$&  116.527   & 3&$8.316\pm0.136$&	$7.823  \pm 0.063$ \Tstrut\\
J095904.3+021516&149.768&2.254&0.025&$8.713\pm0.115$&52.729&$1.908\pm0.248$&  29.397   &3&$8.700\pm0.130$&	$7.985  \pm 0.085$ \Tstrut\\
J100211.2+020118&150.547&2.022&0.021&$8.713\pm0.119$&38.652&$1.936\pm0.174$& 42.384   & 3&$9.749\pm0.087$&	$9.181  \pm 0.098$ \Tstrut\\
J100009.3+024247&150.039&2.713&0.033&$9.472\pm0.068$&59.051&$3.277\pm0.295$&  336.929   &5 &$10.402\pm0.094$&	$9.493  \pm 0.094$ \Tstrut\\	
J100115.2+021823&150.313&2.306&0.028&$9.248\pm0.057$&72.663&$3.100\pm0.279$&  172.713   & 4&$9.567\pm0.086$&	$8.792  \pm 0.092$ \Tstrut\\	
J095720.6+015507&149.336&1.919&0.032&$9.690\pm0.034$&45.168&$4.270\pm0.384$&56.609     & 5&$10.557\pm0.096$&	$10.101  \pm 0.076$\Tstrut\\	
J100143.2+024109&150.430&2.686&0.047&$9.712\pm0.065$&43.408&$3.513\pm0.316$&   82.713  & 4&$10.088\pm0.091$&	$9.433  \pm 0.099$ \Tstrut\\
J100259.0+022035&150.746&2.343&0.044&$9.769\pm0.065$&47.931&$4.705\pm0.423$&  256.795   & 6&$11.156\pm0.102$&	$10.921  \pm 0.101$\Tstrut\\
J100055.2+022344&150.230&2.395&0.044&$9.274\pm0.148$&44.149&$2.441\pm0.220$&  210.507   & 5&$10.804\pm0.098$&	$10.531  \pm 0.083$\Tstrut\\
J095923.2+024137&149.847&2.694&0.048&$8.983\pm0.272$&48.236&$3.276\pm0.295$&  83.175   & 3&$10.359\pm0.094$&	$9.753  \pm 0.105$ \Tstrut\\
J100236.5+014836&150.652&1.810&0.046&$9.725\pm0.043$&56.564&$5.582\pm0.502$&   329.657  & 4&$10.010\pm0.090$&	$9.474  \pm 0.127$ \Tstrut\\
J100117.1+020337&150.321&2.060&0.062&$9.317\pm0.114$&49.847&$3.443\pm0.310$&  138.907   & 3&$10.138\pm0.091$&	$9.505  \pm 0.099$ \Tstrut\\
J100003.9+015253&150.016&1.881&0.065&$9.524\pm0.202$&60.666&$3.562\pm0.321$&  18.422   & 5&$9.852\pm0.089$&	$9.194  \pm 0.091$ \Tstrut\\
J095755.9+022608&149.483&2.436&0.071&$9.508\pm0.197$&78.558&$6.341\pm0.571$& 59.042    & 5&$10.664\pm0.097$&	$10.081  \pm 0.104$\Tstrut\\
J100217.9+015124&150.574&1.857&0.062&$9.701\pm0.088$&51.440&$3.743\pm0.337$&  340.661   & 3&$10.779\pm0.098$&	$10.312  \pm 0.122$\Tstrut\\
J100103.7+023053&150.265&2.515&0.072&$9.837\pm0.106$&58.864&$6.061\pm0.545$&  83.189    & 5&$10.824\pm0.098$&	$10.093  \pm 0.105$\Tstrut\\
J095907.8+024213&149.782&2.704&0.079&$9.804\pm0.098$&49.830&$8.351\pm0.752$&  295.425   & 4& $10.996\pm0.100$ &$10.814  \pm 0.077$\Tstrut\\
\hline
\end{tabular}
\end{adjustbox}
\caption{Table summarising our sample main properties. Column (1):
Galaxy identifier; Column (2): Right ascension of the galaxy in degrees (J2000); Column (3): Declination of the galaxy in degrees (J2000); Column (4): Redshift derived from the \hi\ line; Column (5): Log$_{10}$ of the galaxy’s \hi\ mass; Column (6): Inclination angle (in degrees), measured from IRAC 1 imaging; Column (7): Stellar disc scale length in kiloparsecs; Column (8): Position angle (in degrees), used for kinematic modelling;
Column (9): Number of the resolution elements across the major axis of the \hi moment 1 map; Column (10): Log$_{10}$ of total luminosity in the Spitzer 3.6~$\mu$m
band; Column (11): Log$_{10}$ of the galaxy's stellar mass derived using resolved $\Upsilon_{*}$. 
}
\label{tbl_sample}
\end{table*} 

\begin{table*} 
\centering
\begin{tabular}{lccc}   
\hline
\hline\Tstrut
Name& log$_{10}$(M$_{200}^{\Upsilon_{*}=0.5}$) & log$_{10}$(M200$_{200}^{\Upsilon_{*}=0.2}$)& log$_{10}$(M$_{200}^{ \rm SED \Upsilon_{*}}$)\\	
    & M$_{\odot}$ & M$_{\odot}$ & M$_{\odot}$\\
 \hline
J095829.1+014139&10.895$^{+0.10}_{-0.09}$&10.896$^{+0.10}_{-0.09}$& -- \Tstrut\\
J095846.8+022051&10.638$^{+0.10}_{-0.09}$&10.645$^{+0.10}_{-0.09}$&10.632$^{+0.10}_{-0.09}$\Tstrut\\
J095927.9+020025&10.405$^{+0.24}_{-0.18}$&10.404$^{+0.24}_{-0.18}$&10.406$^{+0.24}_{-0.18}$\Tstrut\\
J100005.8+015440&10.201$^{+0.35}_{-0.22}$&10.204$^{+0.34}_{-0.23}$&10.200$^{+0.35}_{-0.22}$\Tstrut\\
J095904.3+021516&10.269$^{+0.17}_{-0.14}$&10.273$^{+0.17}_{-0.14}$&10.260$^{+0.17}_{-0.14}$\Tstrut\\
J100211.2+020118&11.431$^{+0.08}_{-0.07}$&11.553$^{+0.09}_{-0.08}$&11.369$^{+0.31}_{-0.24}$\Tstrut\\
J100009.3+024247&11.098$^{+0.08}_{-0.06}$&11.182$^{+0.08}_{-0.07}$&10.996$^{+0.07}_{-0.05}$\Tstrut\\	
J100115.2+021823&10.780$^{+0.13}_{-0.11}$&10.788$^{+0.13}_{-0.11}$&10.783$^{+0.13}_{-0.11}$\Tstrut\\	
J095720.6+015507&11.166$^{+0.20}_{-0.15}$&11.171$^{+0.19}_{-0.15}$&11.509$^{+0.10}_{-0.08}$\Tstrut\\	
J100143.2+024109&11.578$^{+0.10}_{-0.08}$&11.591$^{+0.09}_{-0.08}$&11.570$^{+0.09}_{-0.08}$\Tstrut\\
J100259.0+022035&11.725$^{+0.12}_{-0.12}$&12.121$^{+0.11}_{-0.09}$&11.925$^{+0.09}_{-0.01}$\Tstrut\\
J100055.2+022344&11.453$^{+0.29}_{-0.31}$&11.788$^{+0.28}_{-0.25}$&11.540$^{+0.18}_{-0.15}$\Tstrut\\
J095923.2+024137&11.588$^{+0.11}_{-0.09}$&11.604$^{+0.12}_{-0.10}$&11.571$^{+0.11}_{-0.09}$\Tstrut\\
J100236.5+014836&10.858$^{+0.09}_{-0.08}$&10.899$^{+0.09}_{-0.08}$&10.848$^{+0.09}_{-0.08}$\Tstrut\\
J100117.1+020337&11.371$^{+0.12}_{-0.10}$&11.385$^{+0.12}_{-0.10}$&11.352$^{+0.12}_{-0.10}$\Tstrut\\
J100003.9+015253&11.299$^{+0.08}_{-0.08}$&11.315$^{+0.08}_{-0.08}$&11.282$^{+0.08}_{-0.07}$\Tstrut\\
J095755.9+022608&11.914$^{+0.09}_{-0.08}$&11.947$^{+0.09}_{-0.08}$&11.803$^{+0.08}_{-0.07}$\Tstrut\\
J100217.9+015124&11.867$^{+0.14}_{-0.11}$&11.888$^{+0.14}_{-0.12}$&11.826$^{+0.13}_{-0.11}$\Tstrut\\
J100103.7+023053&11.281$^{+0.10}_{-0.09}$&11.358$^{+0.10}_{-0.09}$&11.094$^{+0.10}_{-0.09}$\Tstrut\\
J095907.8+024213&12.148$^{+0.12}_{-0.10}$&12.193$^{+0.12}_{-0.11}$&11.978$^{+0.11}_{-0.10}$\Tstrut\\
\hline
\end{tabular}
\caption{Table summarising the DM halo mass (M$_{200}$) obtained using different assumption of the mass-to-light ratio in mass modelling. Column (1):
Galaxy identifier; Column (2): M$_{200}$ assuming $\Upsilon_{*}=0.5$; Column (3): M$_{200}$ assuming $\Upsilon_{*}=0.2$; Column (4): M$_{200}$ assuming resolved $\Upsilon_{*}$}
\label{tbl_MM}
\end{table*} 

\end{document}